\documentclass[aps,prd,reprint,a4paper,showpacs,nofootinbib,superscriptaddress]{revtex4-1}
\usepackage{natbib}

\usepackage{bbm}
\usepackage{amsmath}
\usepackage{graphicx}
\usepackage{float}
\usepackage{amssymb}
\usepackage{subfigure}
\usepackage{amssymb}
\usepackage{hyperref}
\usepackage{epstopdf}

\usepackage[utf8]{inputenc}
\hypersetup{
    colorlinks=true,
    linkcolor=red,
    citecolor=blue,
}
\usepackage{color}

\providecommand{\dif}{\mathrm{d}} \def\d{\dif}
\usepackage{graphicx}
\usepackage{dcolumn}
\usepackage{bm}
\usepackage{color}
\usepackage{enumitem}
\usepackage{amsmath}
\usepackage{amssymb}

\newcommand{\beq}{\begin{equation}}
\newcommand{\eeq}{\end{equation}}
\newcommand{\bea}{\begin{eqnarray}}
\newcommand{\eea}{\end{eqnarray}}
\newcommand{\non}{\nonumber}
\usepackage[T1]{fontenc}
\usepackage{txfonts}
\usepackage{mathrsfs}
\usepackage{latexsym}
\usepackage{epsfig}
\usepackage{epstopdf}
\usepackage{epstopdf}
\usepackage{graphicx}
\usepackage{amssymb}
\usepackage{amsmath}
\usepackage{dcolumn}
\usepackage{bm}
\usepackage{color}
\usepackage{comment}
\usepackage{xcolor}
\usepackage{dsfont}
\usepackage{orcidlink}
\usepackage{float, geometry,lipsum}
\providecommand{\dif}{\mathrm{d}} \def\d{\dif}

\begin{document}
\title{Disformal Kerr Imprints on BHL Accretion: Shock Morphology, PSD Signatures, and Observational QPO Counterparts}

\author{Orhan~Donmez}
\email{orhan.donmez@aum.edu.kw}
\affiliation{College of Engineering and Technology, American University of the Middle East, Egaila 54200, Kuwait}

\author{M. Yousaf}
\email{myousaf.math@gmail.com}
\affiliation{Department of Mathematics, Virtual University of Pakistan, 54-Lawrence Road, Lahore 54000, Pakistan.}

\author{Imtiaz Khan}
\email{ikhanphys1993@gmail.com}
\affiliation{Department of Physics, Zhejiang Normal University, Jinhua, Zhejiang 321004, China}

\author{G. Mustafa}
\email{gmustafa3828@gmail.com}
\affiliation{Department of Physics, Zhejiang Normal University,
Jinhua 321004, China}

\begin{abstract}
In this work, we reveal the effect of the spacetime parameters on the accretion morphology formed through the Bondi-Hoyle-Lyttleton (BHL) mechanism around a slowly rotating disformal Kerr black hole. Thus, we investigate the measurable signatures of these parameters on the hydrodynamical morphology and the timing behavior of the accreting flow. By numerically solving the general relativistic hydrodynamical (GRH) equations, we reveal the behavior of the accreting matter around a black hole with rotation parameter $a=0.3M$. It is shown that even weak disformal deviations from the Kerr solution modify the shock-cone structure, enhance the density in the post-shock region, and produce coherent oscillations in the accretion rate. Strong deviations produce qualitatively different physical mechanisms. These are classic shock cone, mixed shock-cone/spiral structures, burst-like accretion, and warped post-shock flow. Power spectral density analysis with Lorentzian decomposition reveals model-dependent Quasi periodic oscillations (QPOs) like frequencies, allowing the simulated variability to be compared directly with observed black-hole timing signals. The most important key result of this work is that the obtained QPO-like frequencies can be compared with the observed QPOs from black-hole sources with different masses. The Kerr model produces coherent peaks at $42.99\,\mathrm{Hz}$ and $68.13\,\mathrm{Hz}$, and these frequencies are consistent with the high-frequency QPOs observed from the source GRS 1915+105. In the models where the deviations from the Kerr solution are weak, low-frequency QPOs are produced and found to be coherent. These frequencies also fall within the frequency range observed in Galactic black-hole binaries. On the other hand, the models with large deviations from Kerr can be used to explain observational results that are more irregular, broad-band, and contain multiple peaks. In addition, by using inverse-mass scaling in this work, the numerically calculated frequencies are also compared with observations of intermediate-mass and supermassive black holes. In particular, the disformal black-hole models are found to be consistent with the observational results obtained from the sources M82 X-1, NGC 5408 X-1, and RE J1034+396. This comparison also allows the possible black-hole mass range of observed sources to be inferred from the relation between simulated and observed frequencies. As a result, the numerical results obtained in this work show that variations in the disformal spacetime parameters do not only shift the QPO frequencies, but also lead to the formation of different physical mechanisms that can explain different sources. This makes BHL accretion in disformal Kerr geometry a powerful framework for connecting modified-gravity black-hole spacetimes with observable QPO phenomenology.
\end{abstract}

\maketitle

\date{\today}


\section{Introduction}\label{S1}

Black holes (BHs) are singular entities defined by intense gravitational fields in the cosmic realm. Classically, nothing can escape from the event horizon of a BH due to the powerful gravitational forces; instead, it absorbs everything entangled in its vicinity. The rapid progress achieved in gravitational wave detections and BH imaging has provided a powerful observational platform for examining theoretical descriptions of compact astrophysical objects~\cite{abbott2019tests,abbott2021tests}, while first M87 event horizon telescope results VI, the shadow and mass of the central BH studied in \cite{event2019first}. So far, the available observations remain broadly consistent with the predictions of general theory of relativity (GTR), supporting the view that astrophysical BHs are accurately described, to a high degree, by the Kerr spacetime. Nevertheless, in order to fully use the constraining capability of present and next generation observations for studying possible departures from GTR, it is essential to develop a detailed phenomenological understanding of rotating BHs in theories beyond the Einstein gravity scenario. In this direction, one of the key issues is to assess whether current and future observational data can effectively discriminate the Kerr geometry from alternative Kerr like compact entities that may generate closely similar observational features. The photon rings around Kerr and Kerr-like BHs investigated in \cite{johannsen2013photon}, while Carson and his collaborators \cite{carson2020asymptotically} studied asymptotically flat and parametrized BH metric that preserves Kerr symmetries.

A commonly adopted method for investigating such deviations is to introduce phenomenological modifications into the spacetime metric through arbitrary functions or parameters, which are then constrained by observational data, and another possibility is to explore geometrical extensions involving torsion in the underlying spacetime structure~\cite{johannsen2011metric,cardoso2014generic,papadopoulos2018preserving,achour2025black}, while parameterized non-circular deviation from the Kerr paradigm and its observational signatures, extreme mass ratio inspirals and Lense Thirring effect studied in \cite{ghosh2025parameterized}. Although these approaches are useful as preliminary model independent studies, they also have important drawbacks, whereas in particular, the inserted deviations often do not possess a transparent geometrical or theoretical origin. Moreover, such parameterizations test only the selected class of geometries, rather than providing a direct examination of a fundamental modified theory of gravity. Therefore, a more physically motivated, though technically demanding, route is to obtain exact analytical rotating BH solutions within extended theories of gravity and then investigate their physical, geometrical and observational consequences.

The difficulty of this program becomes clearer when one recalls the special role played by the Kerr solution and the challenges involved in extending it beyond GTR. The Kerr BH is regarded as the unique stationary final configuration produced by the collapse of a sufficiently massive rotating star, while gravitational field of a spinning mass as an example of algebraically special metrics studied in \cite{kerr1963gravitational}, provided that the spacetime is asymptotically flat and the event horizon is regular and non degenerate, i.e., axisymmetric BH has only two degrees of freedom~\cite{carter1971axisymmetric}. This remarkable result is summarized by the well known no hair theorem, while within the scenario of GTR, however, some rotating vacuum geometries distinct from Kerr have also been constructed. Two important examples are the Tomimatsu-Sato solutions, presented in~\cite{tomimatsu1972new} where a new exact solution for the gravitational field of a spinning mass was studied and the Kerr-Levi-Civita family~\cite{barrientos2025new}, both of which admit meaningful static limits. The Tomimatsu-Sato class is asymptotically flat, but it avoids the standard uniqueness theorem by introducing higher multipolar deformations of the Kerr geometry. These deformations generally give rise to naked singularities in the region where an event horizon would otherwise be expected, while the Kerr-Levi-Civita family describes a regular rotating configuration embedded in an asymptotically Levi-Civita background~\cite{stephani2009exact}. Obtaining rotating solutions in closed analytical form is a highly non-trivial task. The original derivation of the Kerr metric was deeply connected with the Goldberg-Sachs theorem~\cite{barrientos2025new}, which algebraically relates special spacetimes to the existence of a non-shear-free geodesic principal congruence. It was also influenced by the Robinson-Trautman class of expanding spacetimes that are shear-free and twist-free~\cite{robinson1962some}, while recent studies on compact entities in modified gravity and nonlinear electrodynamics provided important insights into BH physics beyond GTR. As, horizon free configurations  alternative to BH highlight stability and electromagnetic features at high energies studied in \cite{yousaf2025implications}, while accretion processes extensively explored, demonstrating significant modifications in BH environments due to nonlinear electrodynamics and alternative gravity theories \cite{donmez2026relativistic,donmez2026accretion}. Moreover, observational constraints based on epicyclic frequencies and oscillatory dynamics provided further support for studying BH properties and testing theoretical models \cite{yousaf2026dual}.
These compact configurations play a crucial role in testing established gravitational scenarios and in distinguishing between competing gravitational theories, and they remain a central topic of interest in physics. Research on BH shadows, along with numerous related studies, has broadened this line of inquiry across a range of gravitational models \cite{synge1966escape,abdujabbarov2016shadow,yousaf2024fuzzy,singh2026probing} and further research on various compact objects across different gravitational theories also enriched the field \cite{asad2024evolution,channuie2025traversable,bhatti2026dynamics}.

The QPOs \cite{BZ13, BZ14} detected in X-ray emission spectrum from tiny quasars provide valuable information into these binary systems, in which BHs are surrounded by accretion disks generated by their star partners \cite{pasham2019loud,smith2021confrontation,singh2022low}, while a case for a binary BH system revealed via quasi periodic outflows studied in \cite{BZ15} and Pasham et al. \cite{pasham2025using} also continued their systematic investigations using infrared dust echoes to identify dright Quasi periodic eruption sources. As matter approaches the innermost stable circular orbits (ISCOs) within the disk, make a reasonable increment to friction which  generates strong X-ray \cite{BZ16}, while QPOs are essential to examine the physics of accretion disks and BH properties such as mass, spin, and charge. Both spectroscopic approaches, which analyze photon frequency distributions, and timing methods, which investigate time-dependent photon counts, are critical to the examination of these systems \cite{remillard2006x}. Ashraf et al. \cite{ashraf2025thermal} studied the motion of test particles around a Schwarzschild BH surrounded by a Dehnen-type DM halo, analyzed how model parameters affect particle dynamics, while the system was characterized by the mass of the BH, the radius of the halo, and the density of the central halo. Analytical expressions for the particles' energy, angular momentum, effective potential, and ISCOs derived, whereas the epicyclic oscillations near the equatorial plane investigated through radial, vertical, and orbital frequencies, along with periastron precession. Furthermore, particle collisions near the horizon and the resulting center of mass energy evaluated which revealed the strong dependence of particle dynamics on the model parameters, also orbital motion and QPOs testing around rotating Hairy BHs, while extensive studies focused on QPOs in the vicinity of BHs under strong gravitational fields \cite{rezzolla2004new,belloni2012high}.

The present work investigates the hydrodynamical and timing signatures of BHL accretion around a slowly rotating disformal Kerr BH \cite{achour2026circular}, whereas this analysis is motivated by the fact that accretion flows in the strong field region can carry observable imprints of the underlying spacetime geometry. In particular, deviations from the standard Kerr background may not only modify the near horizon flow morphology, but may also alter the time dependent variability of the accretion rate, therefore, the study of BHL accretion in a disformal Kerr geometry provides a useful scenario for connecting modified BH spacetimes with possible observational signatures in X ray binaries, ultraluminous X ray sources, and active galactic nuclei. In order to examine these effects systematically, we consider a slowly rotating disformal Kerr background and compare its accretion behavior with the corresponding Kerr reference case. The disformal model contains conformal and disformal parameters that control the deviation from the standard Kerr geometry while keeping the horizon location fixed for the physically allowed cases considered in this study. This setup enables us to isolate the influence of the disformal modification itself, rather than attributing the changes in the accretion dynamics to variations in the BH horizon.

The numerical part of this work is devoted to solving the general relativistic hydrodynamical equations for supersonic matter falling toward the BH through the BHL mechanism, while the simulations allow us to follow the formation and evolution of the post shock region, including the development of shock cones, spiral like overdensities, warped structures, and non axisymmetric flow patterns. As shown in~\cite{Koyuncu:2014MPLA,Donmez2012MNRAS,Donmez2024MPLA}, the formation and evolution of shock cone structures can give rise to QPO type oscillatory behavior. These studies also revealed that the generated QPO frequencies are not universal, but are affected by the BH spin, the dynamical properties of the accreting flow, and the parameters that encode modifications in the spacetime geometry. These features are diagnosed through two dimensional density maps on the equatorial plane, one dimensional azimuthal profiles, and time dependent density distributions, consequently, in this way, the study provides a direct connection between spacetime parameters and the morphology of the accreting matter. A central objective of the paper is to show how different disformal deviations affect the shock cone structure produced by BHL accretion, however the weak deviation models preserve the general shock cone character, but they broaden, displace, and oscillate the downstream high density region relative to the Kerr case. In contrast, stronger deviations can substantially deform the post shock flow, giving rise to mixed shock cone and spiral like configurations. The azimuthal density and velocity diagnostics provide an additional way to understand the near BH flow structure, and by comparing the Kerr model with the slowly rotating disformal cases, we show that even small deviations from the Kerr background can significantly change the density enhancement, radial inflow behavior, and angular motion of the post shock matter. These effects become particularly important in the strong field region, where the shock front, the opening angle of the cone, and the angular displacement of the high density region are most sensitive to the modified geometry. Beyond the spatial morphology, this work also examines how the modified flow structure is transferred into the time domain through the mass accretion rate. Since the shock cone and spiral like overdense regions continuously modulate the amount of matter crossing the inner boundary, the accretion rate becomes a direct diagnostic of the underlying hydrodynamical variability. The Kerr case produces comparatively weak and nearly steady oscillations, whereas the disformal models exhibit enhanced amplitudes, intermittent bursts, or coherent quasi periodic modulations depending on the strength and direction of the deviation.

To identify the characteristic oscillatory signatures of the simulated accretion flows, we perform power spectral density analyses and decompose the dominant features using Lorentzian components, whereas this procedure allows us to extract QPO like frequencies associated with the shock cone oscillations, warped post shock structures, and spiral density modulations. The Kerr reference model mainly produces relatively stable high frequency components, whereas the weak disformal models shift the timing power toward lower frequencies. Stronger deviations generate either broad multi peak variability or state dependent oscillations, depending on the resulting hydrodynamical configuration. Also, the numerically obtained QPO like frequencies are compared with observational timing features reported in BH systems and by applying the standard mass scaling argument, the simulated frequencies can be related not only to Galactic stellar mass BH binaries, but also to ultraluminous X-ray sources and active galactic nuclei. This comparison indicates that the Kerr reference model is more naturally associated with high frequency QPO behavior, while the weak disformal cases are more suitable for low frequency QPO like variability. The strongly deformed cases, on the other hand, may be relevant for irregular, burst like, or state dependent timing behavior.
This paper is structured as follows: In section~\ref{S2}, we introduce the slowly rotating disformal Kerr BH spacetime and discuss the essential features of the corresponding geometrical background, while section~\ref{Numeric_1} is devoted to the analysis of BHL accretion dynamics and quasi periodic signatures around the slowly rotating disformal Kerr BH. In subsection \ref{Numeric_2}, we investigate the two dimensional rest mass density morphology of the accreting flow, with particular attention to the formation and deformation of shock cone structures under different disformal Kerr models, however the azimuthal density distribution and the time dependent evolution of the flow are examined in subsection~\ref{Numeric_3}. Section~\ref{Numeric_4} presents the timing properties of BHL accretion by analyzing the accretion rate variability and the associated quasi periodic oscillation frequencies, while in subsection~\ref{Numeric_5}, we study the variations of the mass accretion rate for different disformal Kerr configurations as well as discuss how the model parameters influence the temporal behavior of the accreting matter, and the power spectral density analysis used to identify QPO signatures is described in subsection~\ref{Numeric_6}. In section~\ref{Numeric_7}, we provide a mass scaled interpretation of the simulated QPO frequencies and compare them with possible observational counterparts of BHL induced variability in astrophysical BH systems and section~\ref{Concl} summarizes the main findings of this work.

\section{Slowly rotating disformal rotating Kerr black hole}\label{S2}

Motivated by the use of disformal transformations as a fundamental solution-generating method, we introduce here a new Kerr black hole obtained via a disformal transformation. Our construction starts from the Kerr stealth solution. To implement the disformal transformation, which is defined as:
\begin{equation*}
g_{ij} \rightarrow C(\varphi, X) g_{ij}+D(\varphi, X) \partial_i \varphi \partial_\nu \varphi. \tag{1sol1}
\end{equation*}
We begin by evaluating the gradient of the scalar field configuration, which yields
\begin{equation*}
\partial_r \varphi=-r \sqrt{\frac{-2 X_0}{\Delta}}, \quad \partial_\theta \varphi=\sqrt{-2 X_0} a \cos \theta . \tag{1sol2}
\end{equation*}
As an initial step in our analysis, we assume that the conformal and disformal factors take the simple forms $C(\varphi, X)=C_0$ and $D(\varphi, X)=D_0$. For the purposes of the present work, the specific choice of the conformal factor is not essential, as it is typically introduced to map between two given theories, often in conjunction with a redefinition of the scalar field. The central ingredient in our framework is the disformal factor. Although, in principle, it may be an arbitrary function of $\varphi$ and $X$, we restrict our attention, for simplicity, to the case of a constant disformal factor, in accordance with previous studies \cite{Anson:2020trg}. Nonetheless, toward the end of this paper we will also consider a more general functional dependence for the disformal factor, specifically of the form $D=D(\varphi)$, thereby ensuring that we remain within the Horndeski class of theories. The line element describing the geometry of a rotating disformal rotating Kerr BH is given by \cite{achour2026circular}

\bea\non
\d s^2 &=& - C_0 \left(\Omega_1\right) \d{t}^2  + 
\frac{C_0 \left(1-\frac{2 D_{0} X_{0}}{C_0}\right)}{1-\frac{2 M}{r}} \d{r}^2 \\\non &+& C_0 r^2 \left(\Omega_2\right)\d\theta^2  - 4 \frac{ a C_0 M \sin ^2(\theta )}{r}\d t \d \phi \\ &+& 
C_0 r^2 \sin ^2(\theta ) \left(1-\frac{a (4 D_{0} X_{0}) \sin (\theta ) \sqrt{r (r-2 M)}}{C_0-2 D_{0} X_{0}}\right) \d \phi^2,\label{BH}
\eea
where 
\begin{eqnarray*}
\Omega_1&&=\frac{a M (4 D_{0} X_{0}) \sin (\theta ) \sqrt{\frac{r-2 M}{r^5}}}{C_0-2 D_{0} X_{0}}+\left(1-\frac{2 M}{r}\right),\\
\Omega_2&&=1-\frac{\sin (\theta ) \sqrt{r (r-2 M)} (4 a D_{0} X_{0})}{r^2 (C_0-2 D_{0} X_{0})}.
\end{eqnarray*}

The solution is obtained by implementing a disformal transformation on a Kerr stealth black hole. A central distinction from previous constructions is that, for the specific seed configuration considered here, the disformal transformation preserves circularity. Consequently, many of the geometric and physical properties that render the Kerr spacetime particularly appealing are retained. We refer to the resulting geometry as the Circular Disformal Kerr spacetime.

\section{BHL Accretion Dynamics and Quasi-Periodic Signatures Around a Slowly Rotating Disformal Kerr Black Hole}
\label{Numeric_1}
In this section, we reveal the dynamical behavior formed as a result of BHL accretion around a slowly rotating disformal Kerr black hole with rotation parameter $a=0.3M$. Our aim here is to show how the disformal spacetime parameters modify the morphological structure formed around the black hole, the stability state of the matter, and the time-dependent variation of the plasma formed as a result of accretion. By analyzing the rest-mass density on the equatorial plane, the density profiles in the azimuthal direction, the temporal variations of the flow, and the resulting time-dependent variations in the mass accretion rates, we aim to identify possible hydrodynamical signatures of the underlying modified geometry. These signatures are especially important because they may reveal the formation of QPO oscillations in the accretion rate. Thus, it becomes possible to establish how the numerically obtained QPOs are connected with the observed QPOs in black-hole systems.

To reveal the accretion dynamics around the black hole and the effect of the spacetime parameters on the resulting physical mechanisms, we numerically solve the GRH equations \cite{Donmez2004ASS, Donmez2006AMC, Donmez2012MNRAS, Donmez2024MPLA}. Through these numerically solved equations, we investigate the behavior of matter that asymptotically falls supersonically from the upstream region at the outer boundary toward the black hole through the BHL mechanism, together with the physical mechanisms produced by this process. At the same time, we reveal the effect of the spacetime parameters on the resulting physical mechanisms. In our numerical modeling, an outflow boundary condition in the radial direction is used in the other boundary regions outside the region where matter is injected toward the black hole. In this way, matter reaching the outer boundary and the inner boundary is allowed to leave the computational domain and fall toward the black hole. On the other hand, periodic boundary conditions are used in the azimuthal direction. Through the numerical simulations, we reveal the time evolution of the accreting matter and determine whether a shock cone, a spiral overdensity region, a quasi-circular structure, or other non-axisymmetric flow behavior forms around the black hole. Thus, by means of these simulations, we directly study how the spacetime parameters control the formation, deformation, and oscillations of the accretion structure.

To reveal the effects of the spacetime parameters on the resulting physical mechanisms and, consequently, on the observable QPO-like structures, the parameter values given in Table \ref{tab:SKBH_parameters} are modeled. In order to compare the slowly rotating disformal Kerr black-hole solutions, the classical Kerr solution is obtained by modeling the cases $C_0=1$ and $X_0=0$ or $D_0=0$, or the case $X_0=0$ and $D_0=0$. On the other hand, the SKBH models represent different configurations of the slowly rotating disformal Kerr black hole. Here, $C_0$ is the conformal parameter that rescales the background metric, $X_0$ denotes the constant background value of the scalar-field kinetic term, and $D_0$ represents the disformal coupling strength. The combination $C_0-2D_0X_0$ is an important expression that controls the regularity and Lorentzian signature of the spacetime. The value of this expression must always be positive in order to obtain a physical black hole. When $C_0-2D_0X_0=1$, the solution becomes the classical Kerr solution, while values larger or smaller than unity indicate the strength of the deviation from the Kerr solution. For all parameter cases given in Table \ref{tab:SKBH_parameters}, the black-hole horizon is located at $r_+=2M$. This allows us to reveal the effects of the disformal parameters on the accretion dynamics, rather than the effects that could arise from a change in the black-hole horizon. By comparing the results obtained from the SKBH1, SKBH2, SKBH4, and SKBH5 models with the Kerr solution, we systematically reveal how variations in expression $C_0-2D_0X_0$ affect the resulting shock-cone structure, density variations, flow asymmetry, and time-dependent changes in the accretion rate.

\begin{table}[htbp]
\centering
\caption{
Model parameters used for BHL accretion simulations around a slowly rotating disformal Kerr black hole on the equatorial plane.  The conformal parameter $C_0$ rescales the background metric, $X_0$ denotes the constant background value of the scalar-field kinetic term, and $D_0$ is the disformal coupling strength.  The combination $C_0 - 2D_0X_0$ controls the regularity and Lorentzian signature of the spacetime and must remain positive to ensure a physical black hole geometry.  The event-horizon radius $r_h$ is computed for $M=1$ and spin parameter $a=0.3M$, and remains fixed at $r_h=2$ for all physically allowed parameter sets.
}
\setlength{\tabcolsep}{12pt}
\label{tab:SKBH_parameters}
\begin{tabular}{cccccc}
\hline\hline
Model & $C_0$ & $X_0$ & $D_0$ & $C_0 - 2D_0X_0$ \\
\hline
Kerr  & 1.00 & 0.0 & 0.0   & 1.0 \\
SKBH1 & 1.00 & 0.02 & 0.20 & 0.992 \\
SKBH2 & 1.02 & 0.01 & 0.30 & 1.014 \\
SKBH4 & 0.80 & 0.01 & 0.50 & 0.790 \\
SKBH5 & 1.50 & 0.02 & 0.50 & 1.480 \\
\hline\hline
\end{tabular}
\end{table}

\subsection{Two-Dimensional Rest-Mass Density Morphology of the BHL Flow}
\label{Numeric_2}

In this section, we discuss the morphological structure formed as a result of the accretion of matter around a slowly rotating disformal Kerr black hole. In previous BHL accretion studies around Schwarzschild and Kerr black holes, it has been shown that matter falling supersonically from the upstream region toward the black hole is gravitationally focused and accumulates on the opposite side of the black hole, namely in the downstream region, producing a dense and well-defined shock region. The resulting cone is generally characterized by a compressed high-density wake, the formation of a clear opening angle, and an oscillatory post-shock motion. In our previous studies \cite{Koyuncu:2014MPLA, Donmez2012MNRAS, Donmez2024MPLA}, it was shown that shock-cone dynamics produces QPO oscillations, and that the resulting QPO frequencies depend on the rotation parameter of the black hole, the flow parameters, and the modification parameters of the spacetime. Similar behavior has also been revealed in different gravity theories where the spacetime is modified. In these studies, it was shown that the modified spacetime can significantly change the morphology of the shock cone, the density distributions, the formation of different physical mechanisms, and the strength of the oscillations in the mass accretion rate \cite{donmez2026relativistic, Al-Badawi:2025yqu, Donmez2024Universe, Donmez2025EPJC, Mustafa2025JCAP, orh1_1}.

\begin{figure*}[tbhp]
\centering
\includegraphics[width=5.5cm,height=5.0cm]{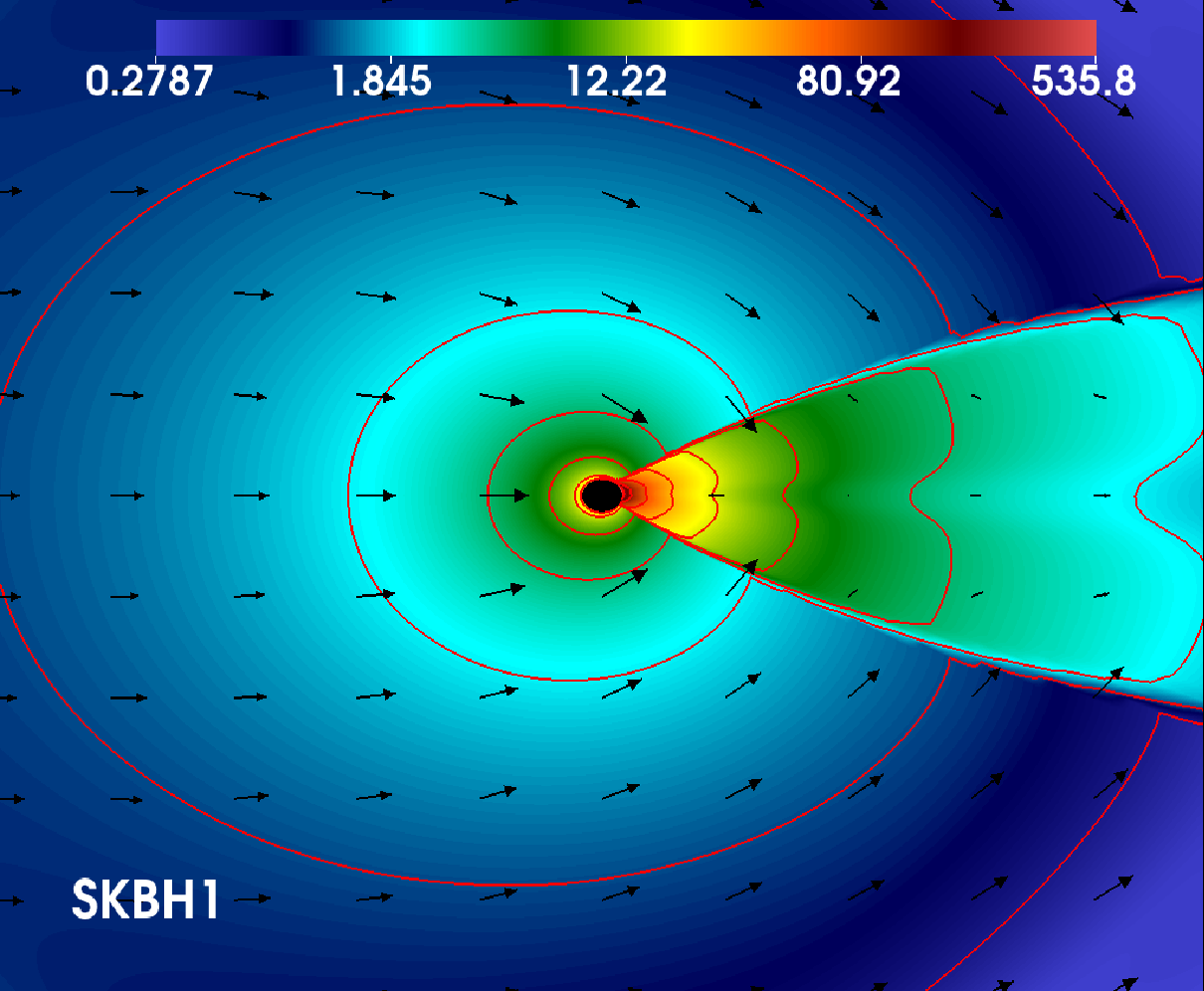}
\includegraphics[width=5.5cm,height=5.0cm]{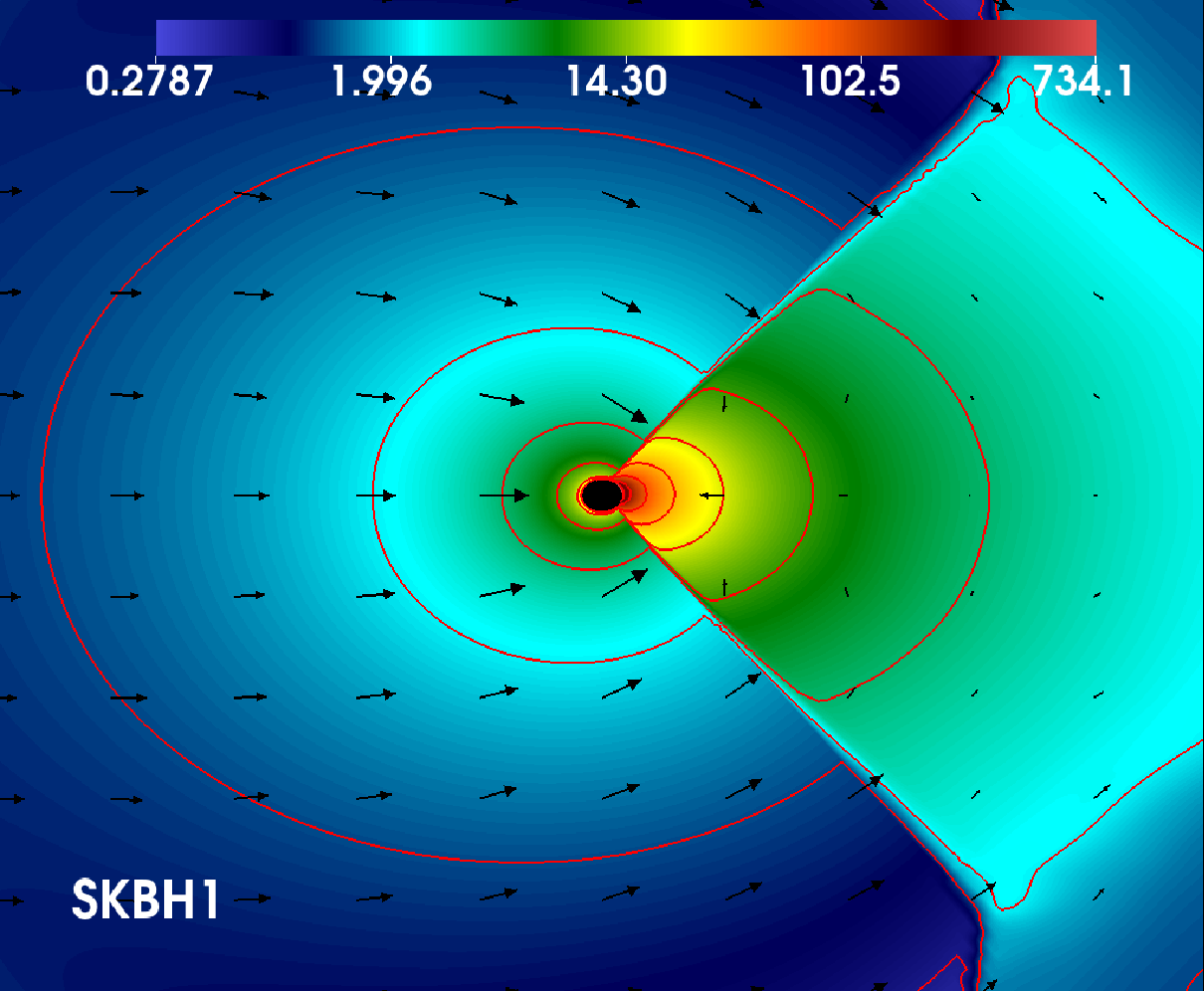}
\includegraphics[width=5.5cm,height=5.0cm]{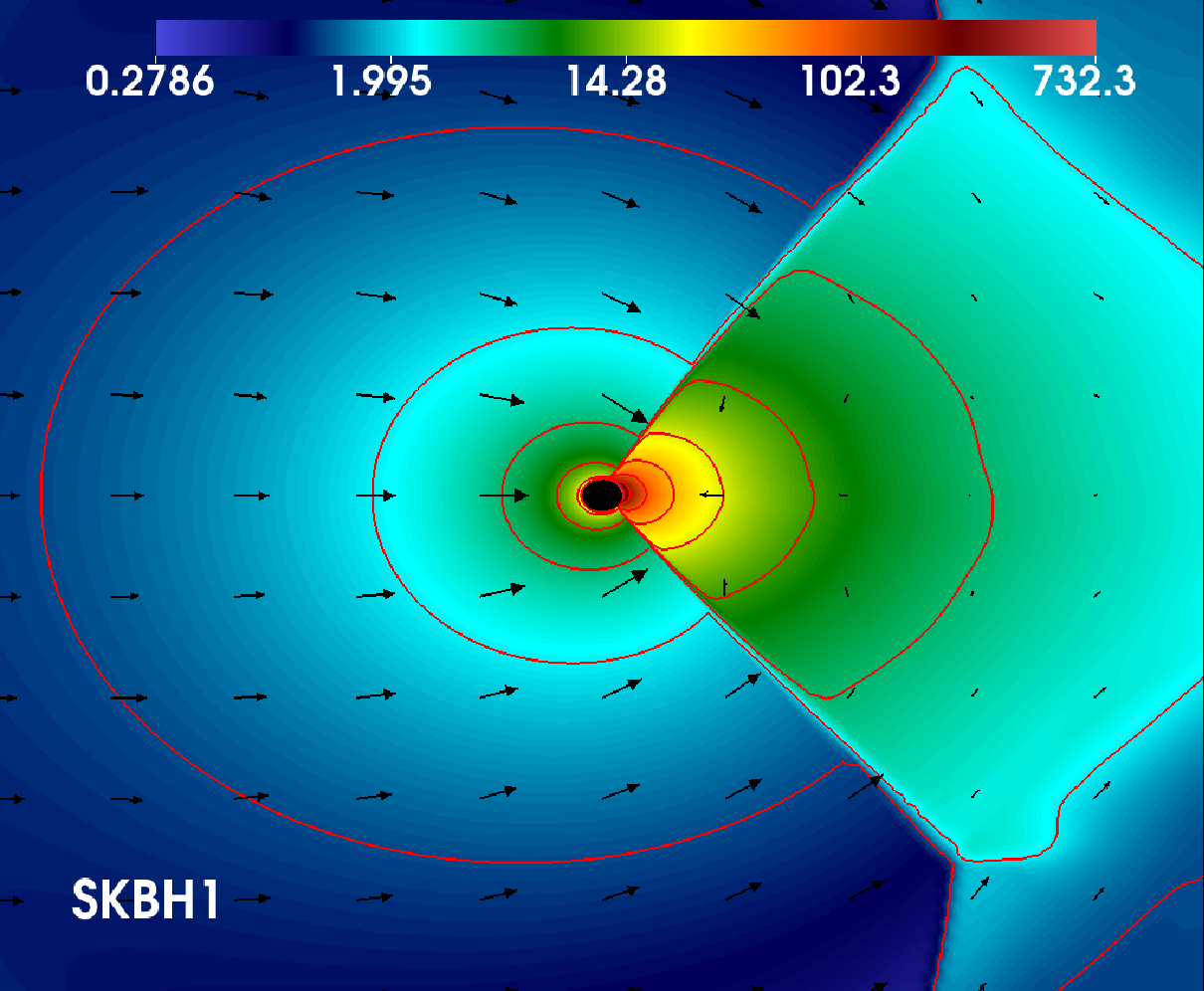}\\
\includegraphics[width=5.5cm,height=5.0cm]{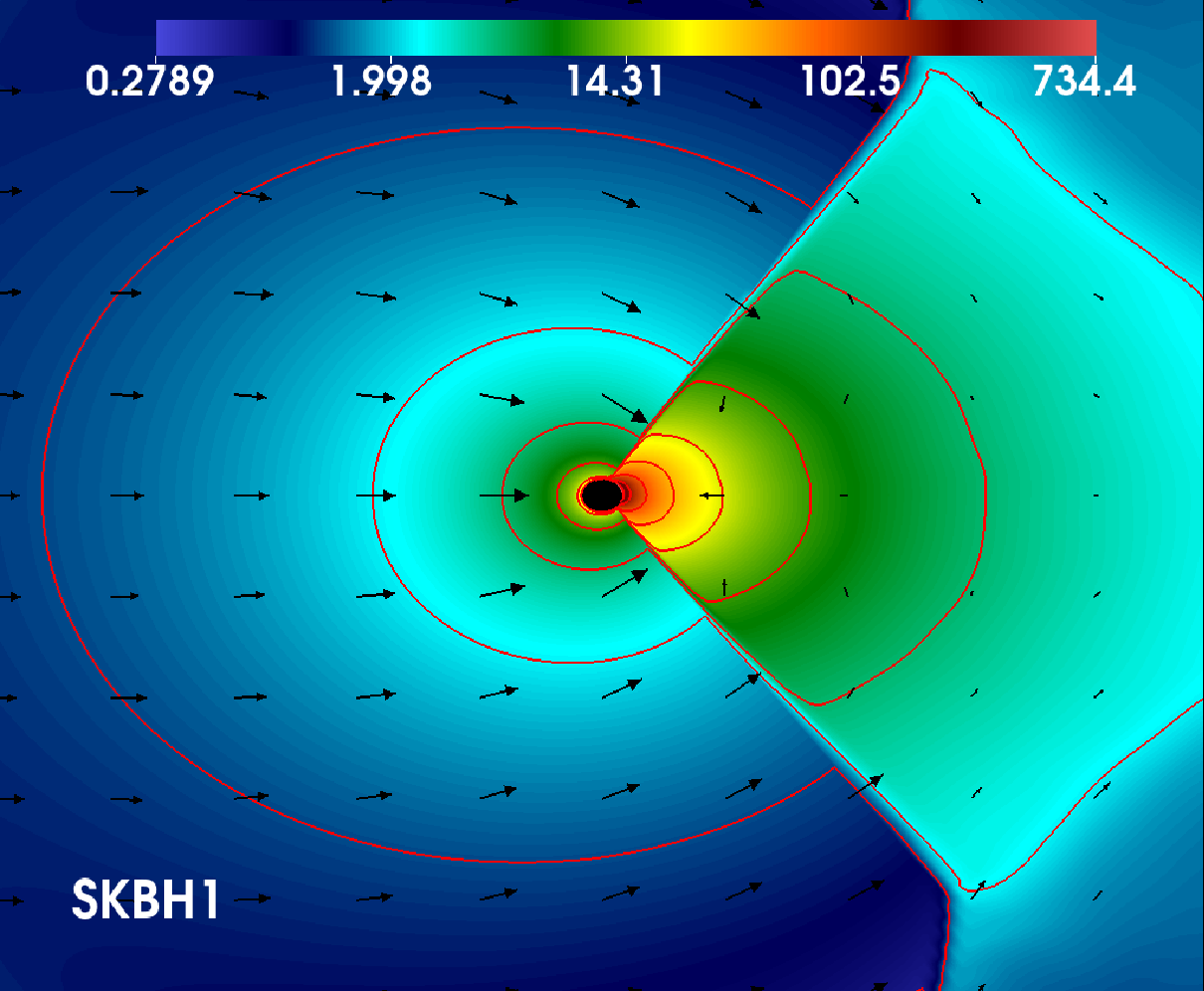}
\includegraphics[width=5.5cm,height=5.0cm]{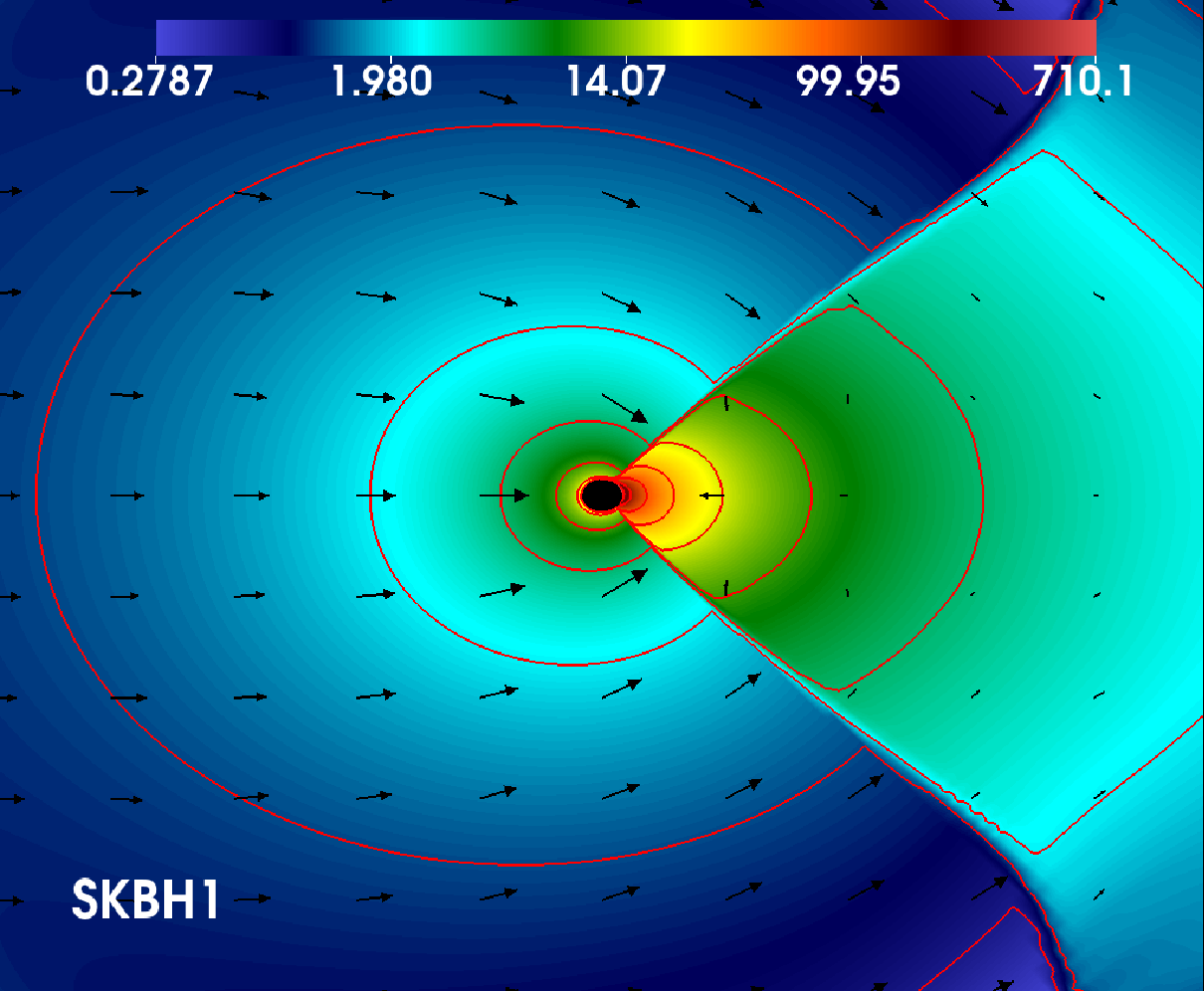}
\includegraphics[width=5.5cm,height=5.0cm]{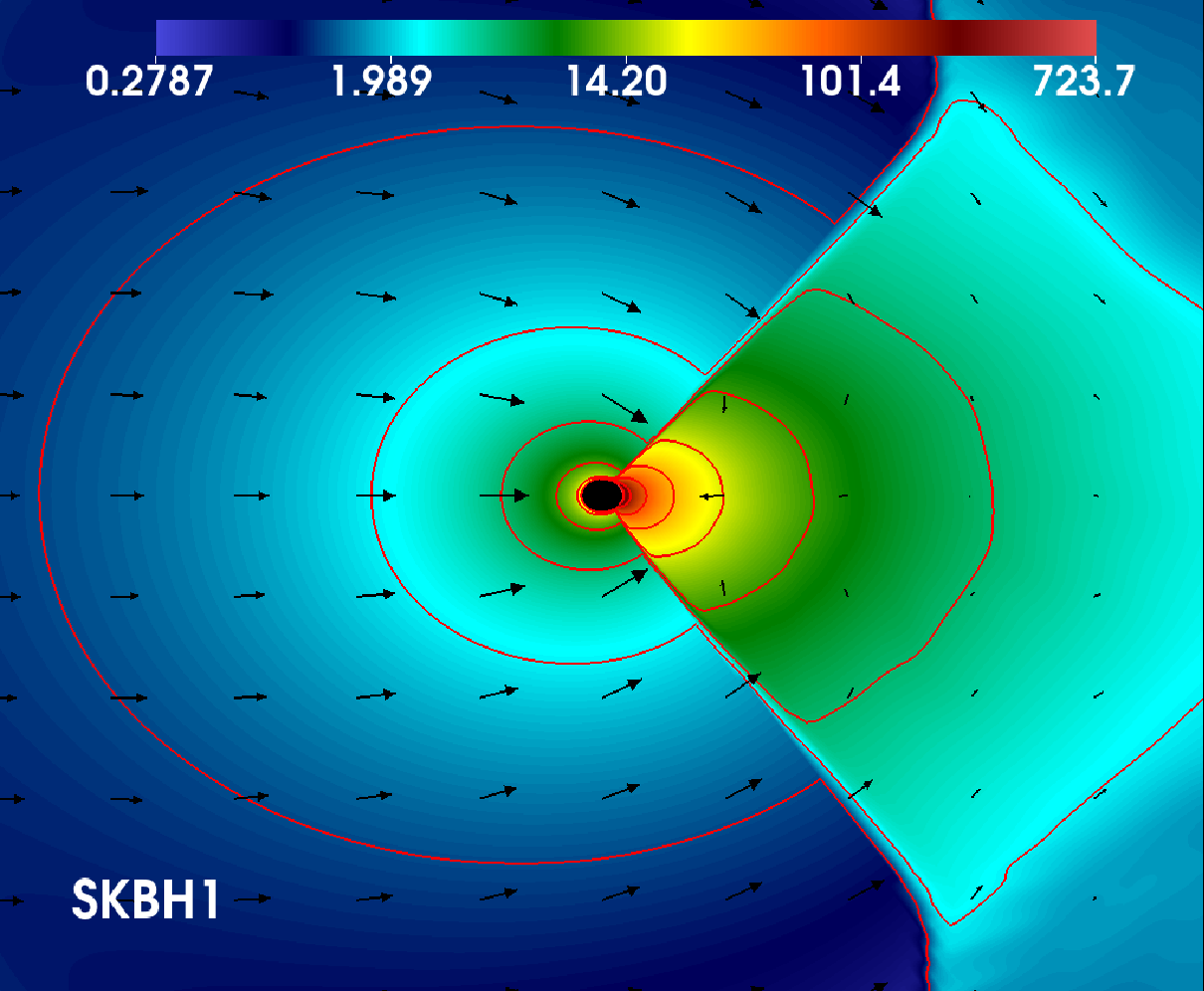}
\caption{Time evolution of the rest-mass density distribution on the equatorial plane for the SKBH1 model of the slowly rotating disformal Kerr black hole. The density is represented by the color scale and contour lines, while the velocity vector field is overlaid to show the direction of the accreting matter and the flow behavior after the shock cone starts to oscillate. The panels show the dynamical evolution of the BHL flow from the upper-left to the lower-right panel. To resolve the near-black-hole accretion morphology and the downstream shock-cone structure in more detail, the computational domain is zoomed into the region $-70M < x < 70M$ and $-70M < y < 70M$.}\label{color_SKBH1}
\end{figure*}

Fig.\ref{color_SKBH1} shows the time evolution of the rest-mass density formed as a result of accretion around a slowly rotating disformal Kerr black hole for the SKBH1 model given in Table \ref{tab:SKBH_parameters} on the equatorial plane. As time progresses, the evolution of the resulting morphology proceeds from the upper-left panel to the lower-right panel. In this way, the time-dependent change of the morphology is presented. In the early stages of the simulation, shown in the upper-left panel, a shock cone begins to form in the downstream region. Although it is not as well developed as in the classical Schwarzschild and Kerr solutions, the formation of the cone can still be observed. Different from the classical Kerr solution, in the SKBH1 model the opening angle of the cone increases with time, and a cone that covers a wider region is formed. As time evolves, the high-density shock-cone region extends further in the azimuthal direction. Although the opening angle of the cone becomes wider, the density near the black hole is observed to increase further. However, the wake formed in the downstream region does not appear at a fixed position. Instead, it starts to oscillate strongly in the azimuthal direction, and this oscillation is clearly seen in Fig.\ref{color_SKBH1} through the snapshots taken at different times. This shows that the formed cone is not a stationary structure as a whole. Rather, as a result of the interaction between the disformal spacetime geometry and the hydrodynamical effects of the matter falling supersonically toward the black hole, the shape, width, and orientation of the dynamically evolving post-shock region are continuously modified.

For the SKBH1 model in Fig.\ref{color_SKBH1}, the disformal parameters are $C_0=1.00$, $X_0=0.02$, and $D_0=0.20$. From these values, $C_0-2D_0X_0=0.992$ is obtained. This value is slightly smaller than that of the classical Kerr solution. Thus, this model represents a weak but non-negligible deviation from the Kerr solution. In both cases, the horizon location remains the same. Since the horizon location is unchanged, the differences seen in the shock morphology are mainly associated with the modification of the spacetime structure rather than a change in the black-hole size. The slight decrease in the value of $C_0-2D_0X_0$ modifies the geometrical response of the flow and causes a broader and less sharply collimated structure in the downstream region. The spreading of the matter trapped inside the shock over a wider region in the azimuthal direction, together with the time-dependent change in the shock opening angle, shows that the disformal correction strengthens the non-axisymmetric deformation in the BHL accretion flow. This result is also physically important because such an oscillating shock cone can periodically change the mass accretion rate and naturally lead to observable QPO-like peaks.

\begin{figure*}[tbhp]
\centering
\includegraphics[width=5.5cm,height=5.0cm]{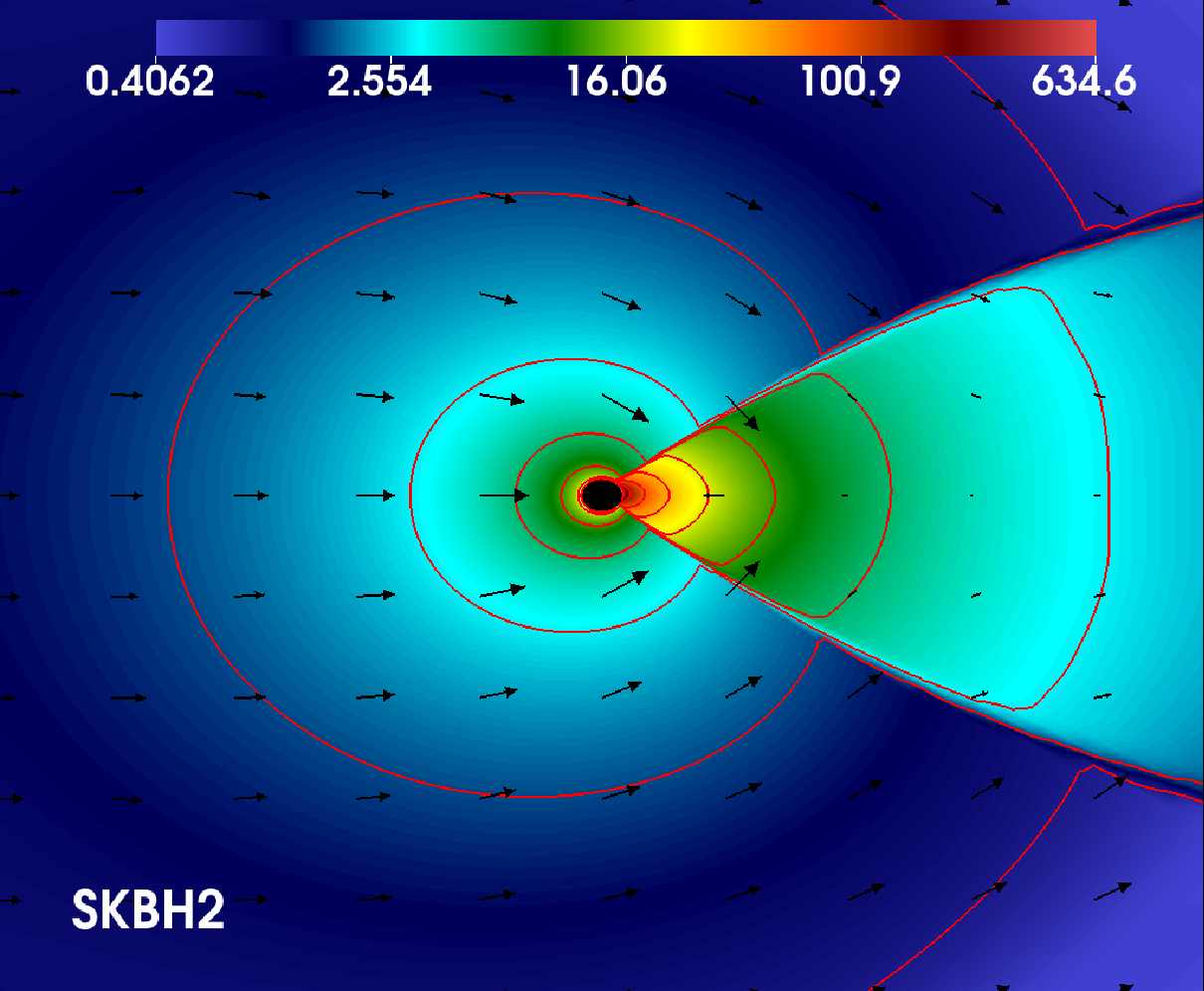}
\includegraphics[width=5.5cm,height=5.0cm]{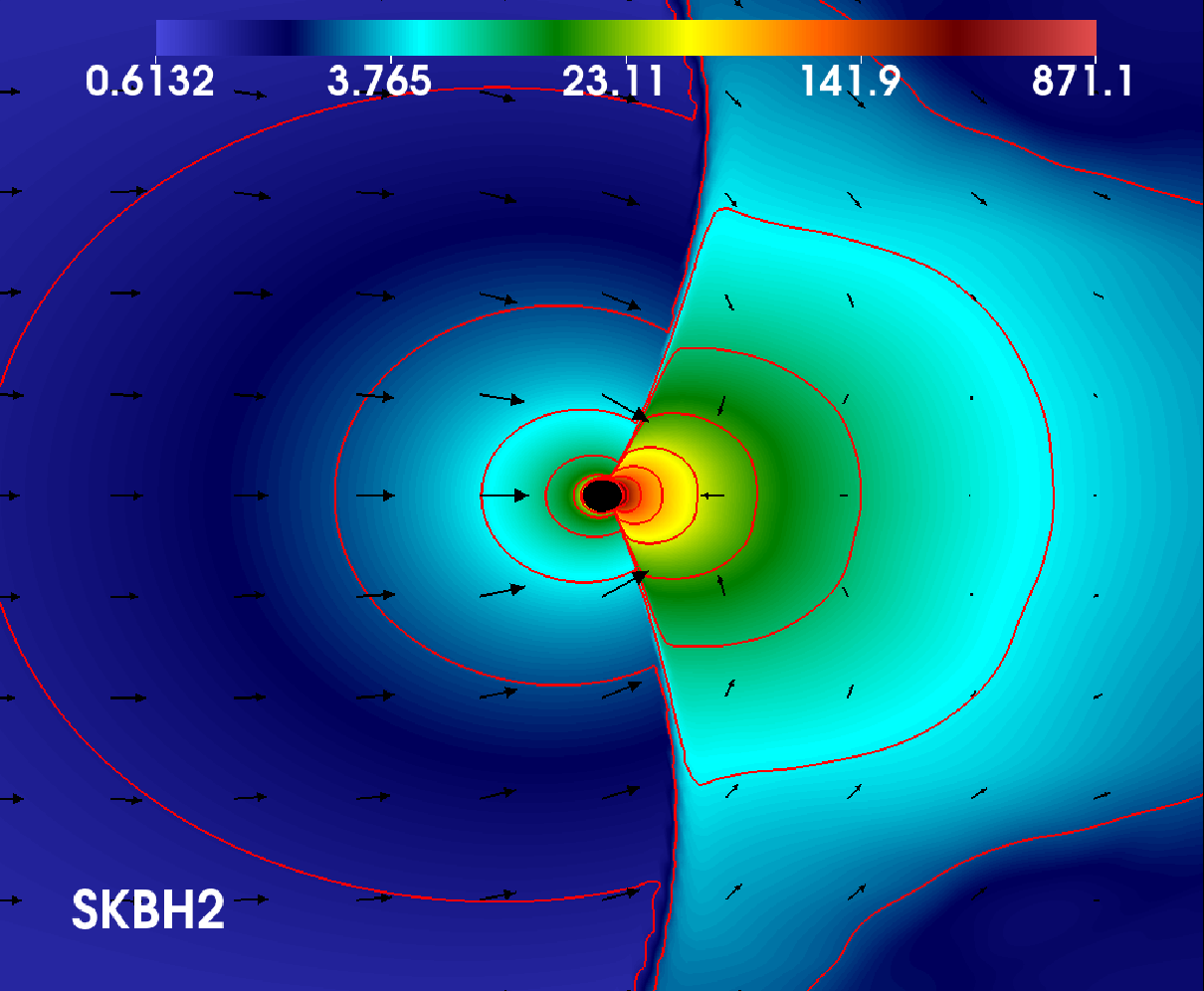}
\includegraphics[width=5.5cm,height=5.0cm]{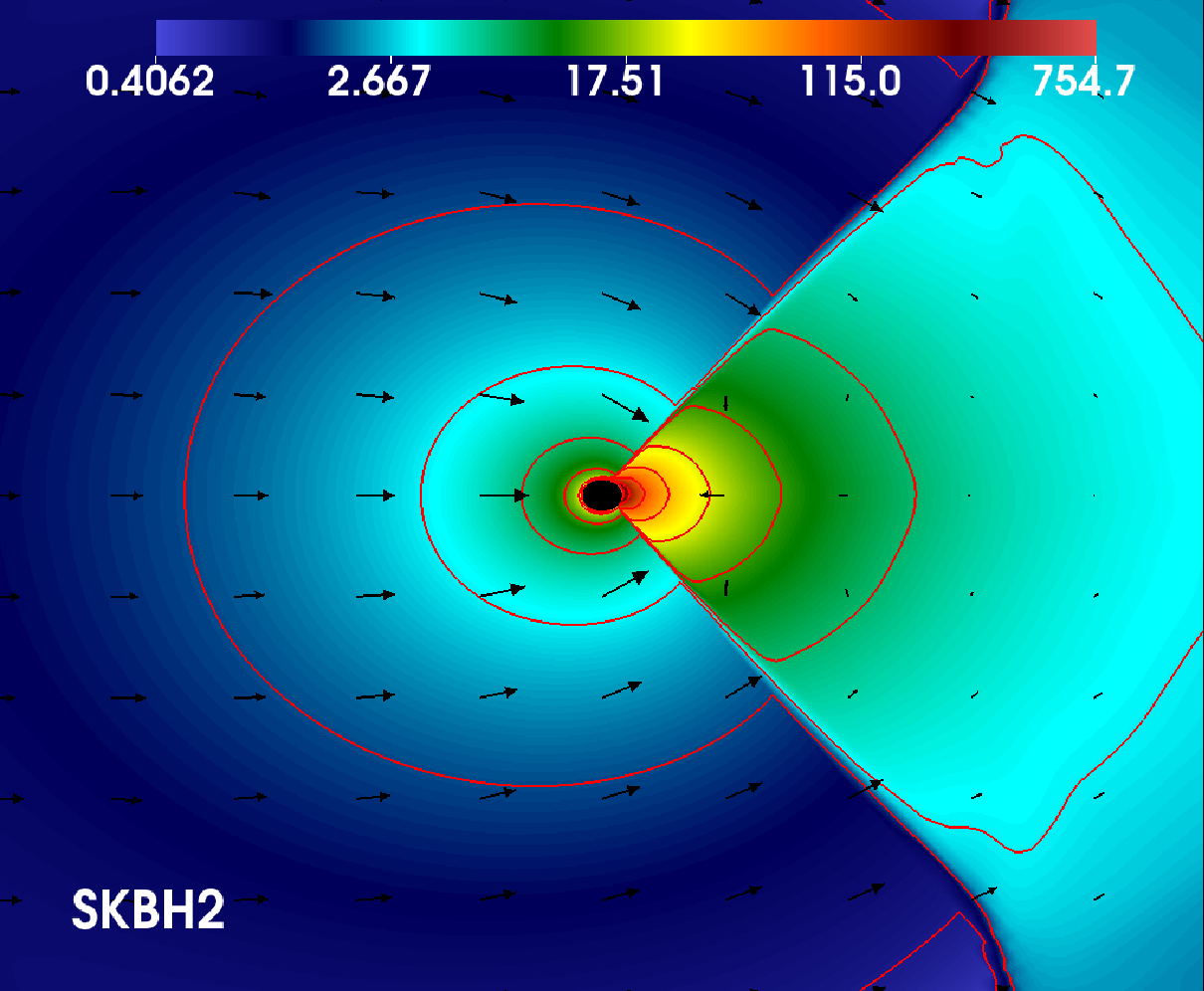}\\
\includegraphics[width=5.5cm,height=5.0cm]{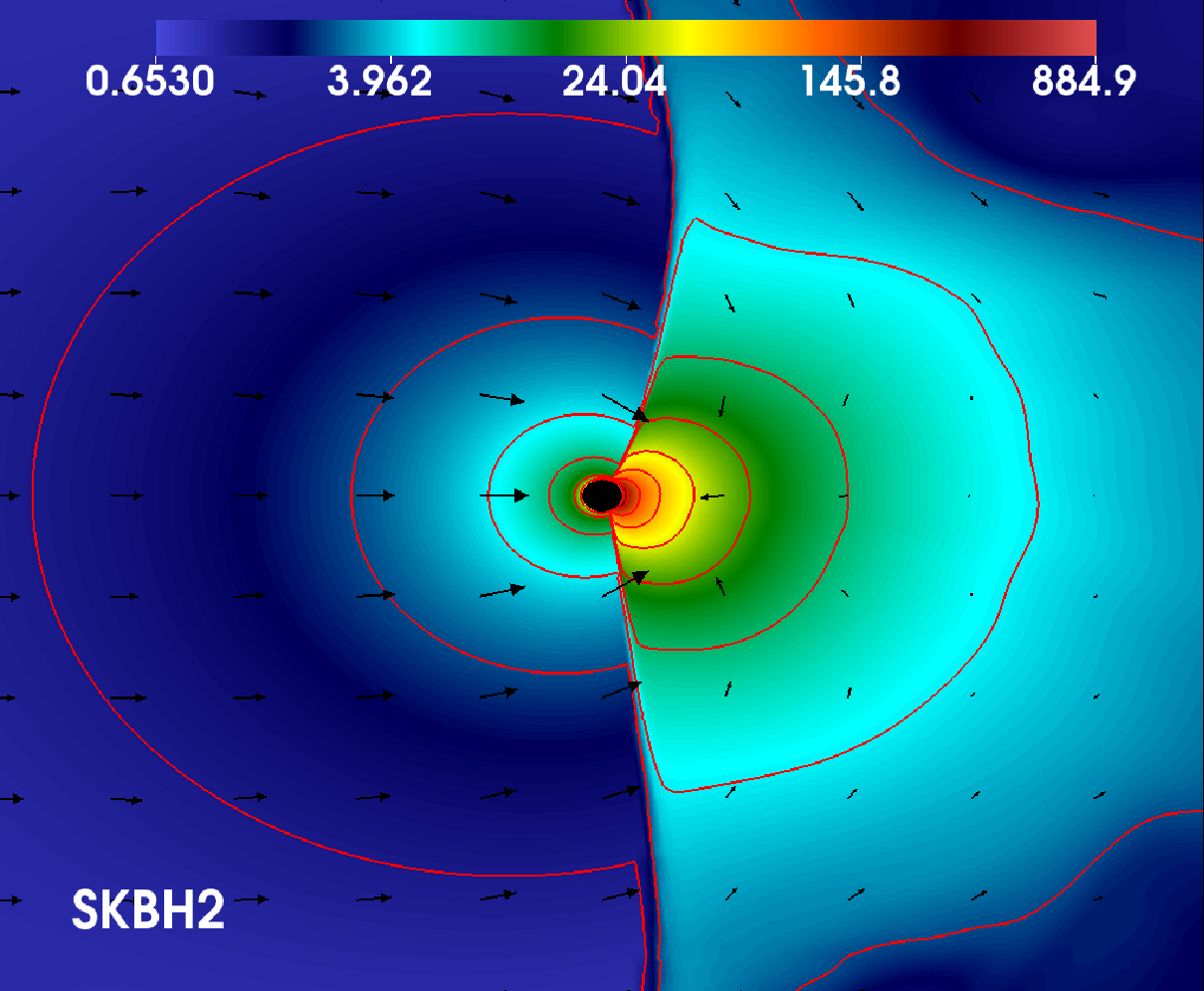}
\includegraphics[width=5.5cm,height=5.0cm]{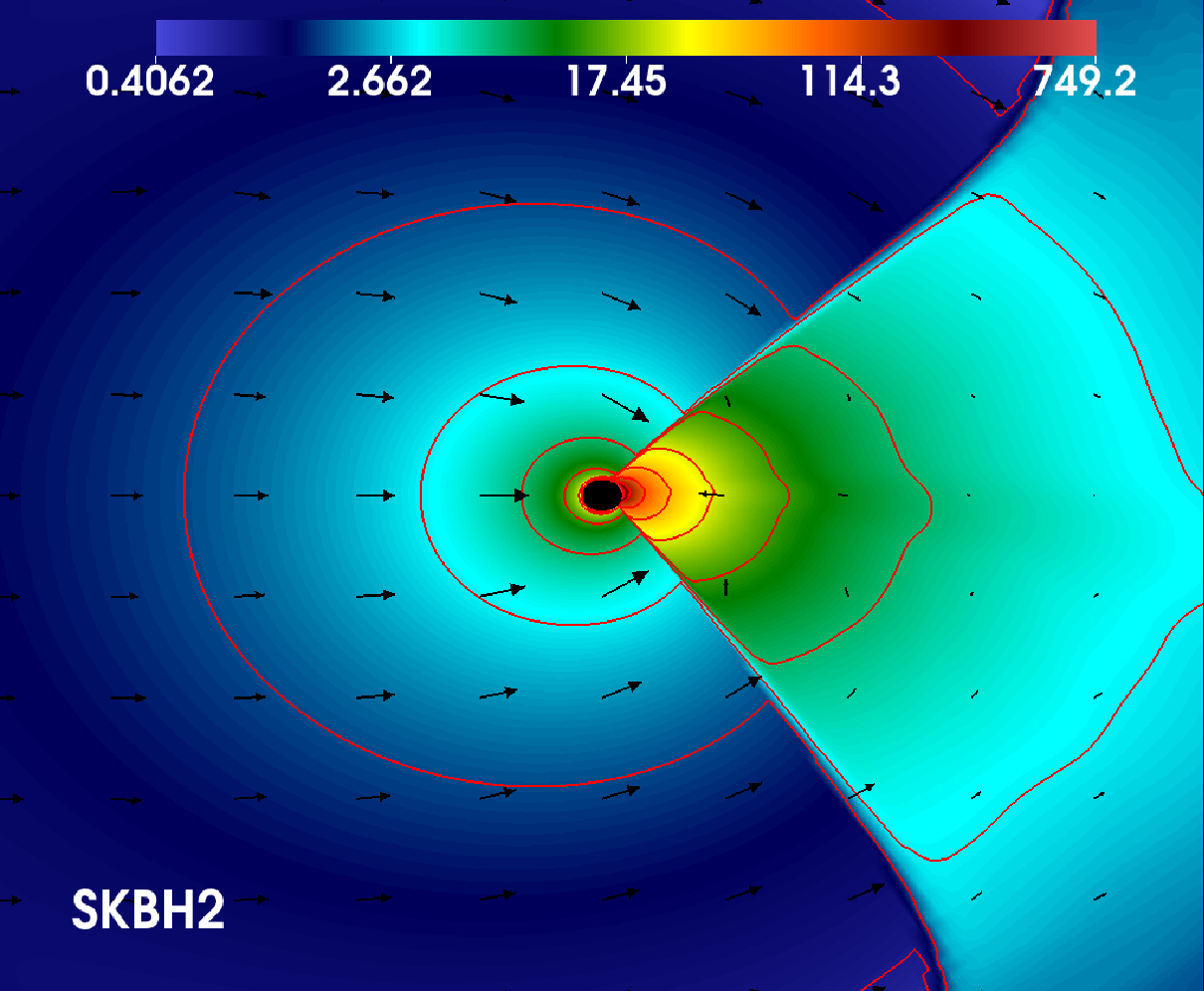}
\includegraphics[width=5.5cm,height=5.0cm]{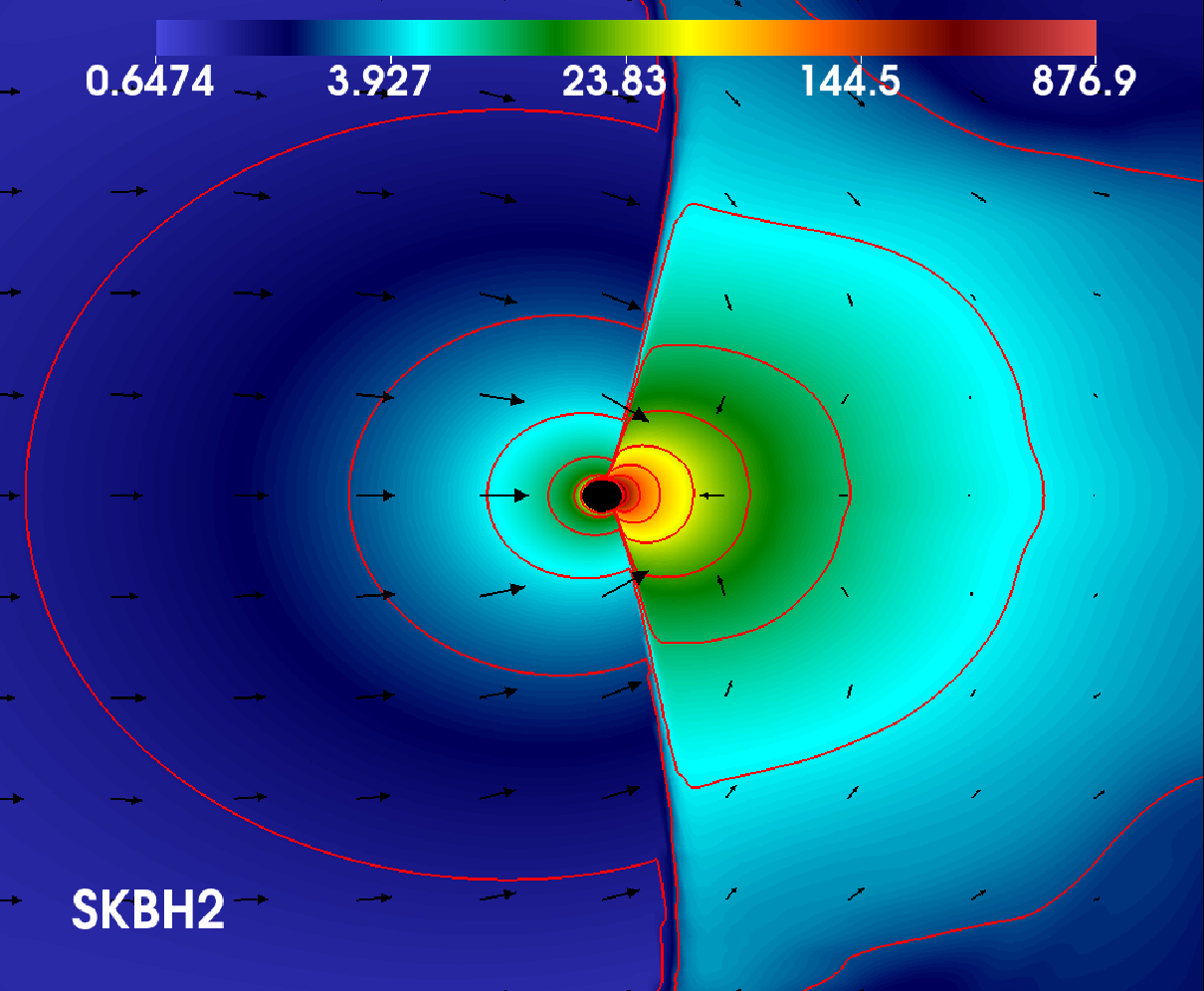}
\caption{Same as Fig.\ref{color_SKBH1}, but for the SKBH2 model given in Table \ref{tab:SKBH_parameters}. In this case, the disformal combination is $C_0-2D_0X_0=1.014$, which is slightly larger than the classical Kerr value. This weak positive deviation from the Kerr solution modifies the shock-cone morphology, producing a more asymmetric and time-dependent downstream structure.}\label{color_SKBH2}
\end{figure*}

The SKBH2 model given in Fig.\ref{color_SKBH2}, similar to Fig.\ref{color_SKBH1}, shows the time evolution of the rest-mass density on the equatorial plane around the slowly rotating disformal Kerr black hole. Similar to Fig.\ref{color_SKBH1} and to the classical Kerr solution, in the early stages of the simulation, the matter falling supersonically toward the black hole is gravitationally focused toward the opposite side of the black hole, leading to the formation of a shock cone. In the SKBH2 model, the spacetime parameters are $C_0=1.02$, $X_0=0.01$, and $D_0=0.30$, while $C_0-2D_0X_0=1.014$. This value is slightly larger than that of the classical Kerr solution. Thus, the SKBH2 model produces a weak positive deviation from the classical Kerr solution. These changes modify the geometrical response of the accreting flow. In this case, the resulting shock cone is still clearly observed throughout the evolution; however, it becomes more asymmetric, and the time-dependent change in the shock-cone structure is clearly seen in Fig.\ref{color_SKBH2}. In the early stage of the simulation, the cone formed in the downstream region appears with a relatively narrow opening angle, similar to the Kerr solution. However, with time, the shock cone spreads over a wider region in the azimuthal direction, and its opening angle changes from one snapshot to another. This shows that the resulting shock cone does not settle into a stable configuration, but instead the entire shock-cone structure undergoes a strong oscillatory motion. Compared with Fig.\ref{color_SKBH1}, the SKBH2 morphology shows stronger changes in the lateral position of the shock boundary, a more pronounced deformation of the downstream high-density region, and larger fluctuations in the density enhancement near the black hole. In particular, the panels show that the shock front exhibits a behavior that alternates between a more collimated shock-cone-like structure and a broader structure. This indicates that even a small positive deviation from the classical Kerr solution can cause significant changes in the morphology of the shock cone by modifying the oscillatory motion, compression, and angular spreading of the matter accreting toward the black hole. Since the horizon remains fixed at $r_+=2M$, these differences are not caused by a change in the black-hole size but by the disformal modification of the spacetime geometry itself. The numerical results of the SKBH2 model show that the morphology of the shock cone formed by BHL accretion can undergo significant changes even for small deviations from the standard Kerr background. These changes strengthen the non-axisymmetric behavior and significantly modify the time-dependent evolution of the accretion flow in the downstream region.

\begin{figure*}[tbhp]
\centering
\includegraphics[width=5.5cm,height=5.0cm]{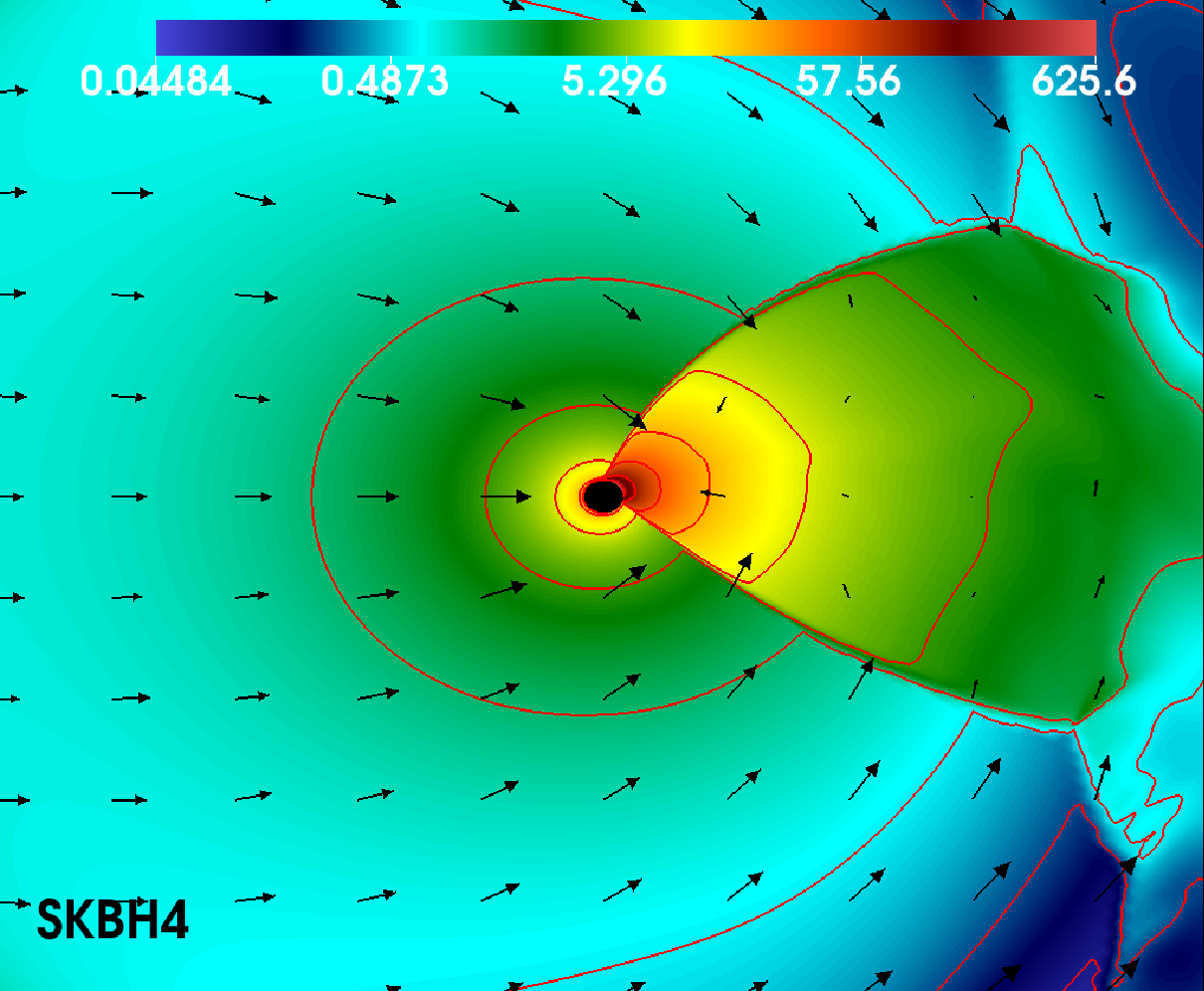}
\includegraphics[width=5.5cm,height=5.0cm]{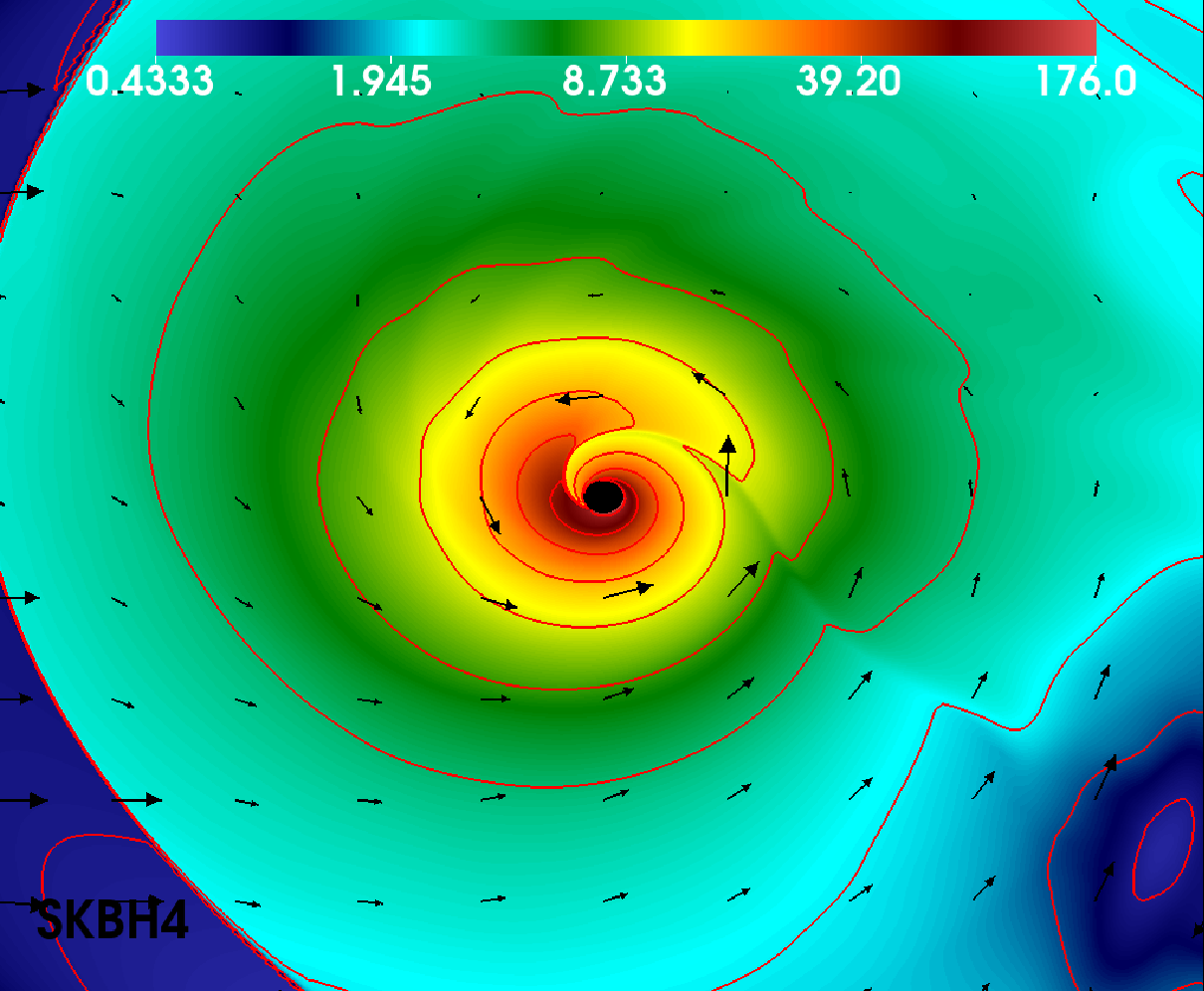}
\includegraphics[width=5.5cm,height=5.0cm]{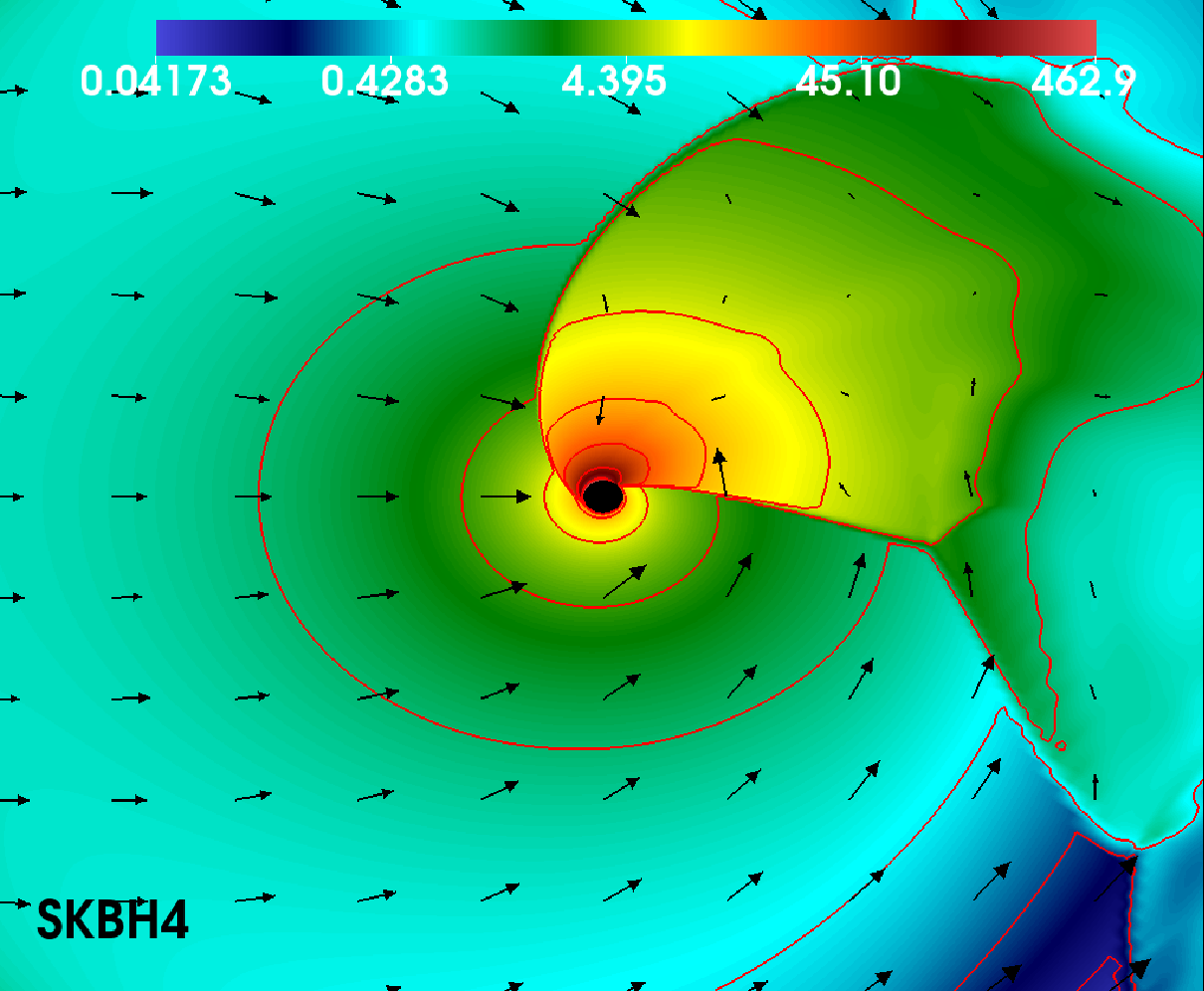}\\
\includegraphics[width=5.5cm,height=5.0cm]{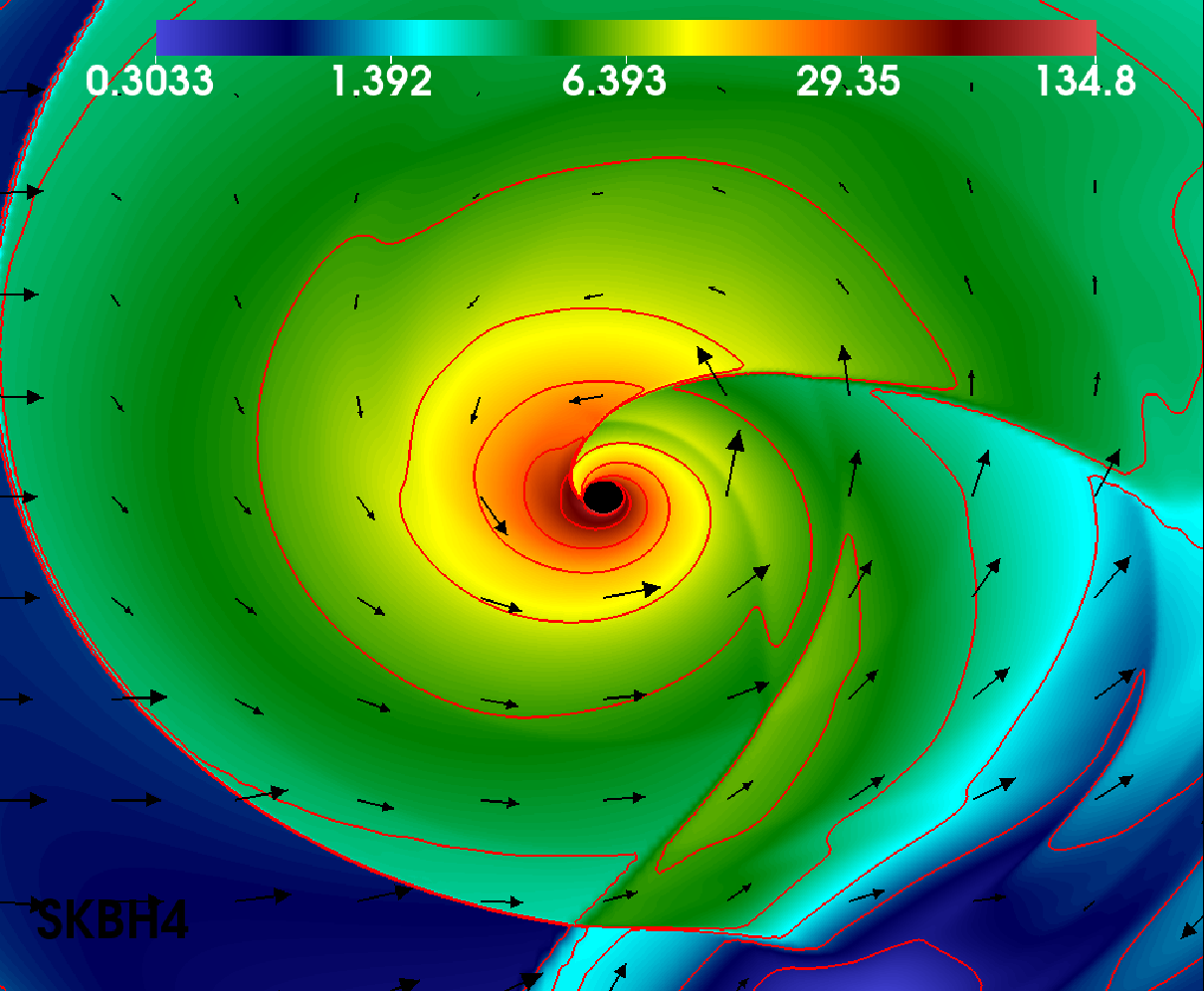}
\includegraphics[width=5.5cm,height=5.0cm]{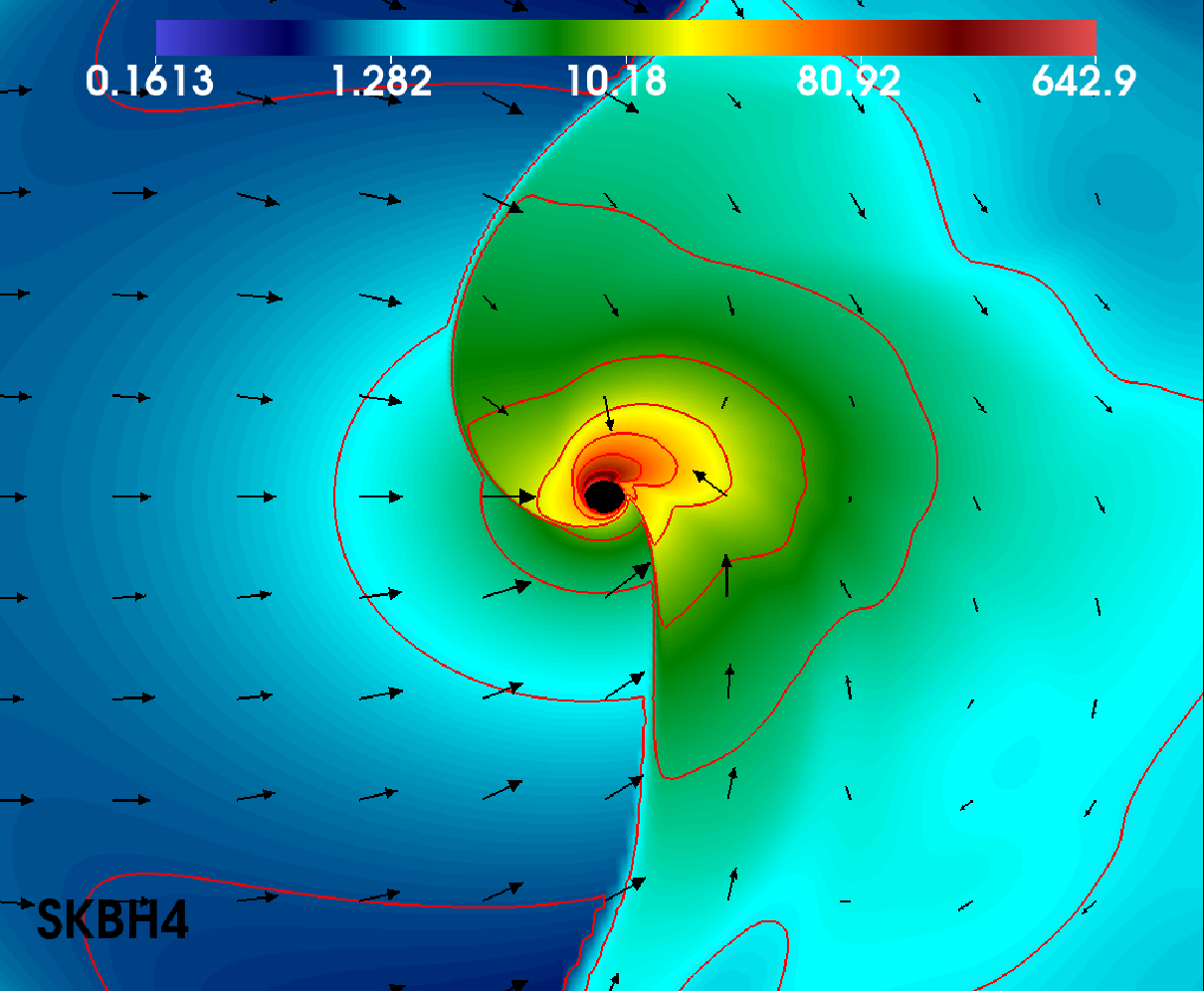}
\includegraphics[width=5.5cm,height=5.0cm]{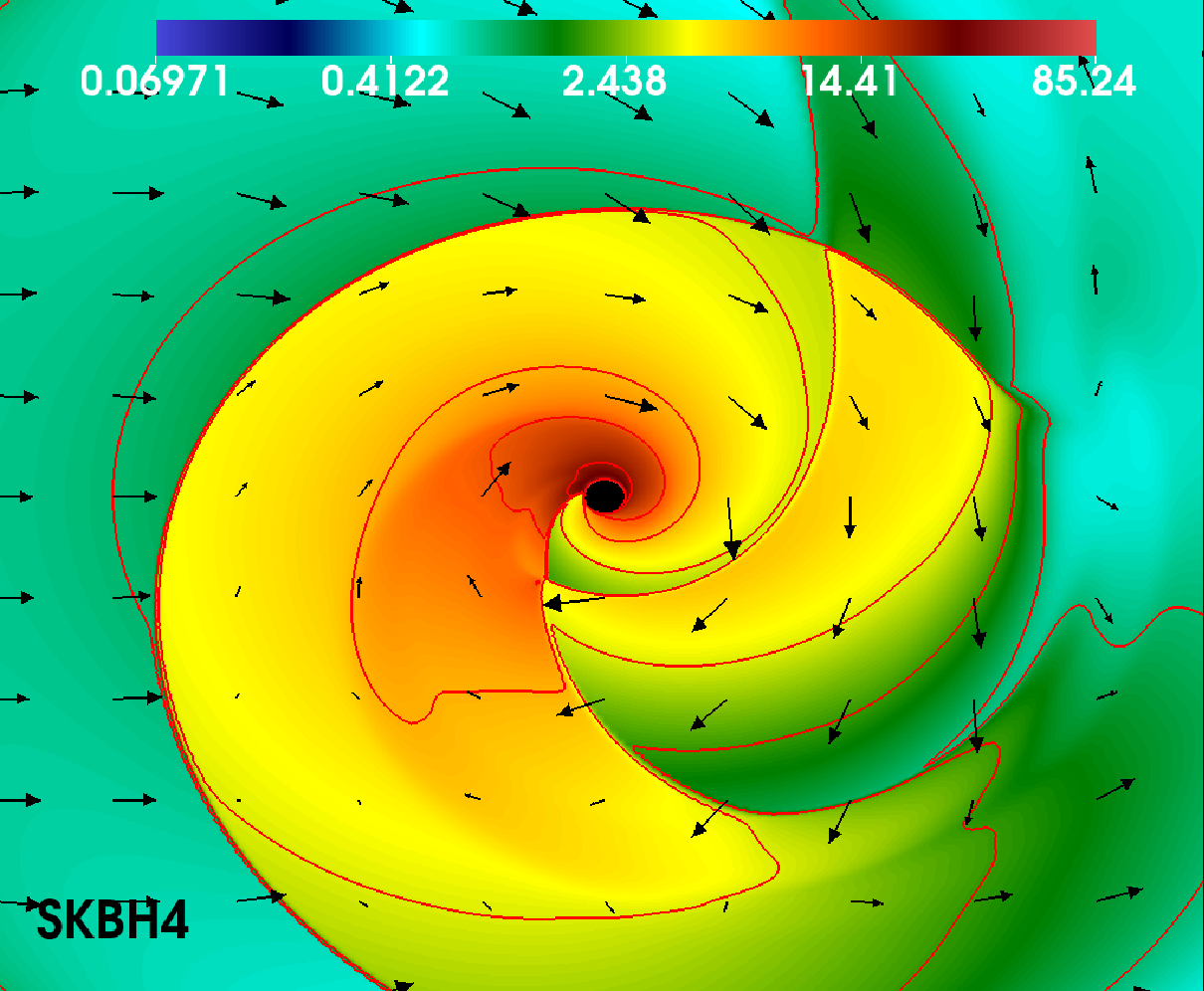}
\caption{Same as Fig.\ref{color_SKBH1}, but for the SKBH4 model given in Table \ref{tab:SKBH_parameters}. In this case, the disformal combination is $C_0-2D_0X_0=0.790$, which is significantly smaller than the classical Kerr value. This strong negative deviation from the Kerr solution strongly modifies the downstream shock morphology, producing a highly non-axisymmetric and time-dependent mixed shock-cone/spiral structure.}\label{color_SKBH4}
\end{figure*}

In the SKBH4 model given in Fig.\ref{color_SKBH4}, the morphological structure formed by the matter accreting toward the black hole is substantially different from the weak-deviation models SKBH1 and SKBH2. In this model, the disformal parameters are $C_0=0.80$, $X_0=0.01$, and $D_0=0.50$, and depending on these parameters, $C_0-2D_0X_0=0.790$ is obtained. This value is significantly smaller than the Kerr reference value $C_0-2D_0X_0=1$. Thus, the SKBH4 model represents a strong negative deviation from the standard Kerr geometry. As a result of this deviation, a narrow and well-developed shock cone does not form in the downstream region as a result of BHL accretion. Instead, as seen in the snapshots in Fig.\ref{color_SKBH4}, a more distorted and dynamically unstable accretion behavior is observed. At the early times of the simulation, a high-density shock region forms in the downstream region. However, the shock-cone morphology seen in the SKBH1 and SKBH2 models rapidly disappears. The matter accumulated in the downstream region spreads over a wider area in the azimuthal direction and causes the formation of a spiral-like dense region around the black hole. In addition, as seen in Fig.\ref{color_SKBH4}, as time progresses, the regions where the matter is dense become bent around the black hole. This shows that the post-shock flow is strongly affected by the modified gravity geometry.

One of the most important features of the SKBH4 model given in Fig.\ref{color_SKBH4} is that the resulting shock structure is strongly non-axisymmetric and strongly time dependent. The shock front does not oscillate only by changing its opening angle, as in the weak-deviation models, but also undergoes a significant global deformation. In other words, the dense regions formed around the black hole rotate around the black hole, surround the black hole, and especially lead to the formation of spiral-like shock structures. The velocity vectors show that the matter is not simply advected downstream after being shocked, but is redirected into a circulating pattern around the black hole. These behaviors show that strong negative deviations in the term $C_0-2D_0X_0$ appear as a physical situation that can explain the transition between the dominant classical BHL shock-cone configuration and the mixed shock-cone/spiral high-density regions formed around the black hole, which can be used to interpret observable data. Thus, the SKBH4 model shows that when the deviation from the Kerr background is sufficiently large, the accretion morphology, compression, and angular momentum transport around the black hole undergo significant changes. Since the black-hole horizon in these models is formed at $2M$, this shows that the change in the physical mechanism observed here does not originate from the black-hole horizon, but is completely a consequence of the disformal modification of the spacetime. The SKBH4 model clearly shows that the spacetime parameters not only modify the density enhancement, but also lead to the formation of a non-axisymmetric hydrodynamical structure. This new hydrodynamical structure appears as a mechanism that can cause the mass accretion rate to vary with time and may lead to the observation of different QPO-like frequencies from the same source.

\begin{figure*}[tbhp]
\centering
\includegraphics[width=5.5cm,height=5.0cm]{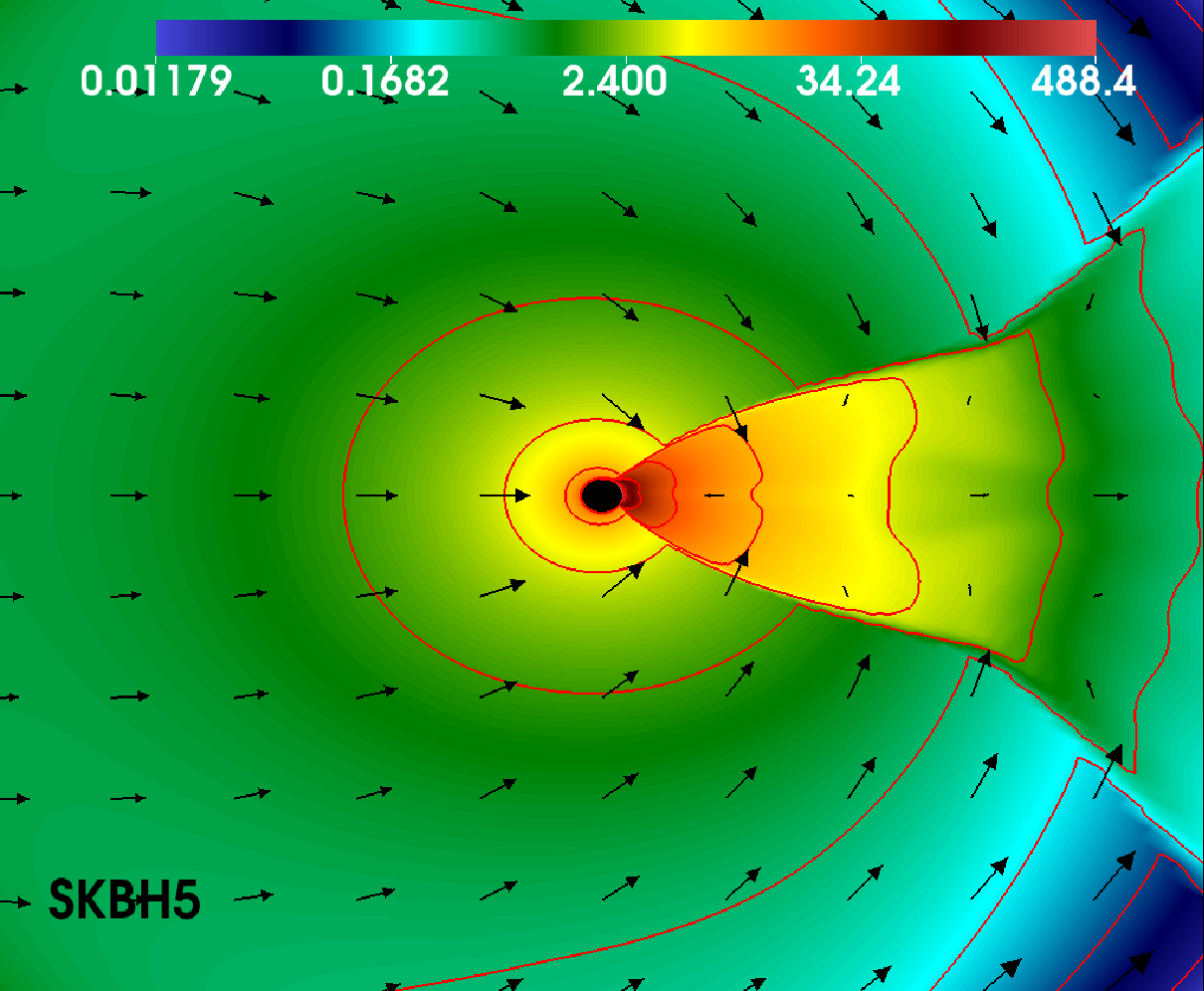}
\includegraphics[width=5.5cm,height=5.0cm]{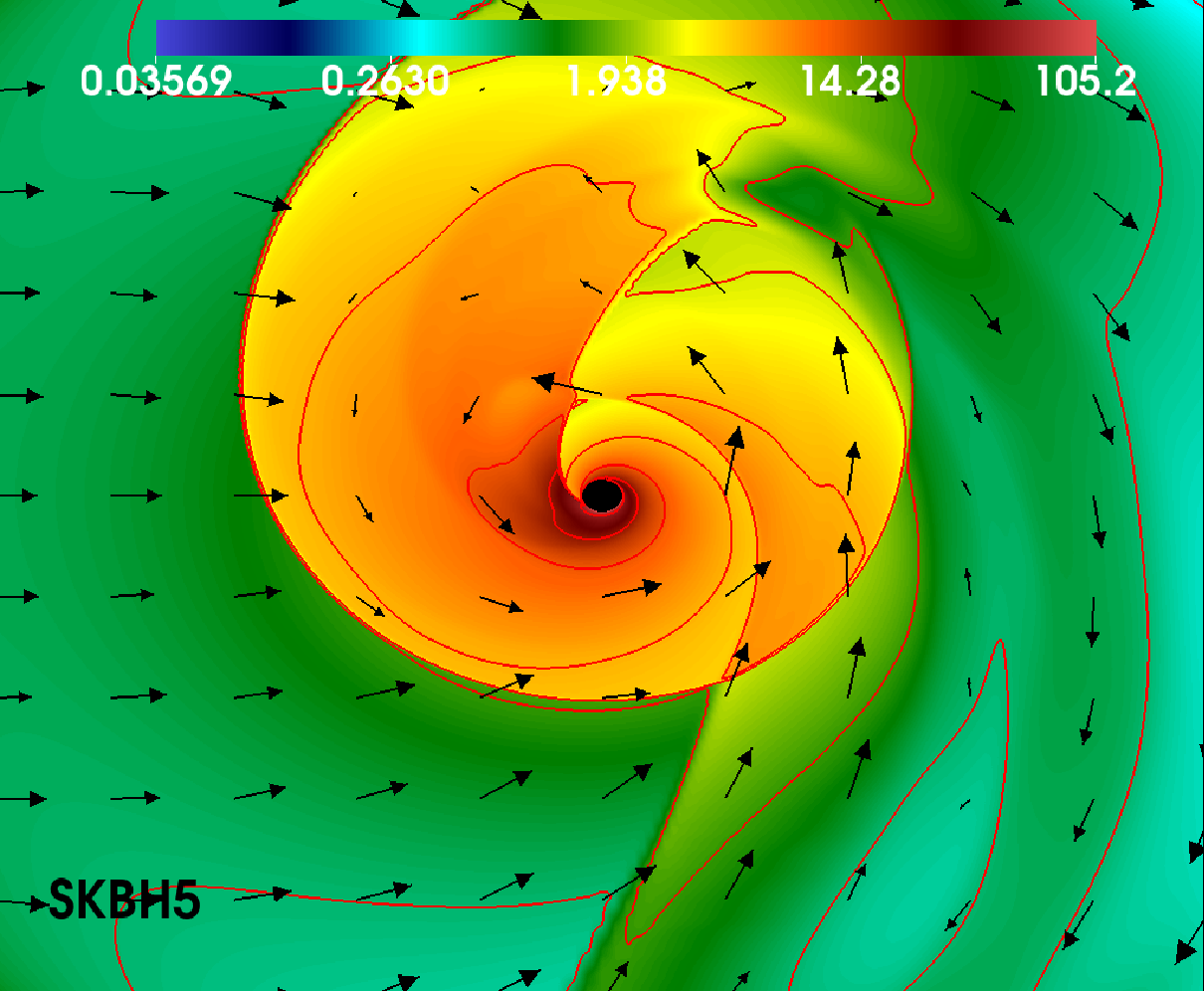}
\includegraphics[width=5.5cm,height=5.0cm]{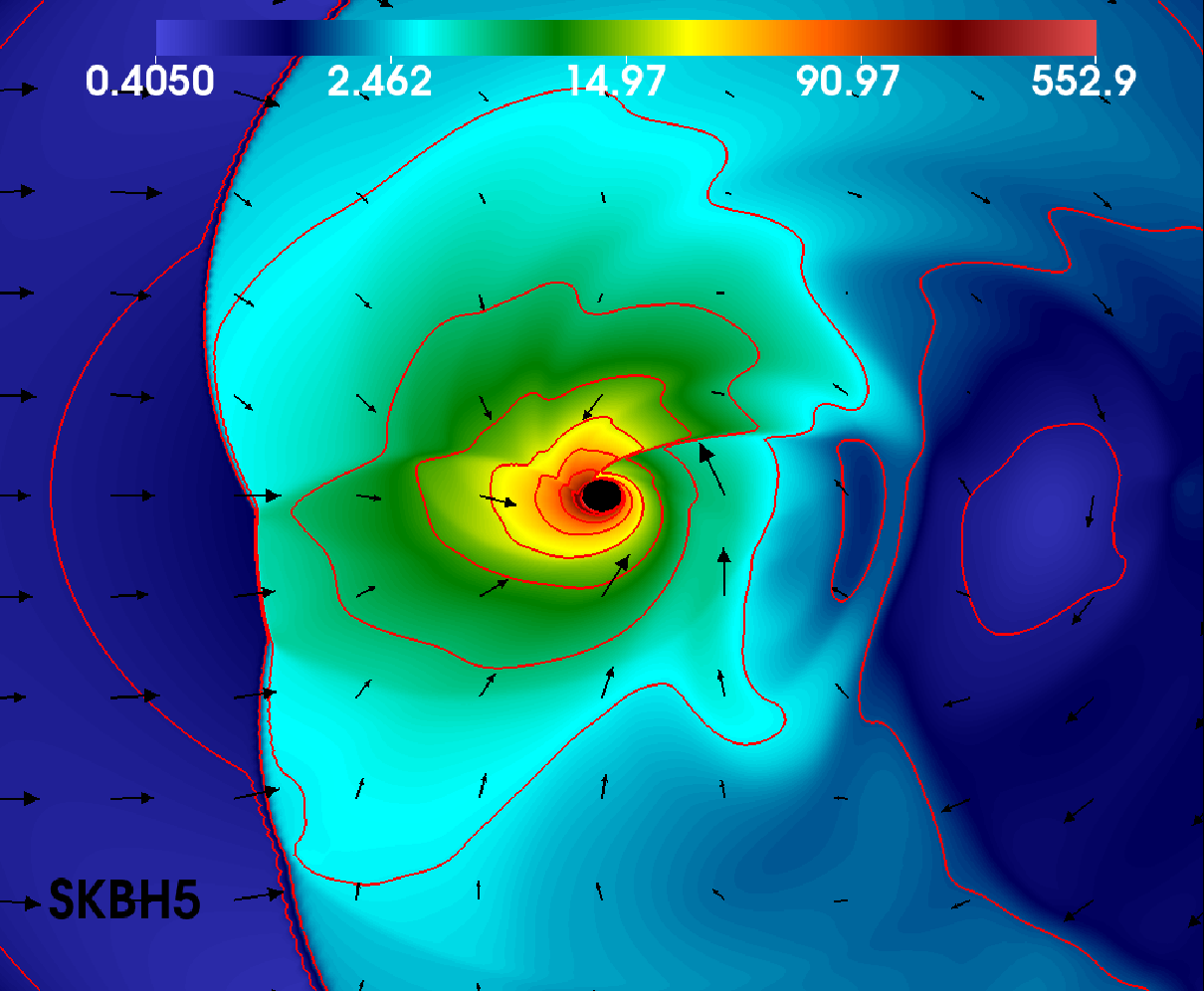}\\
\includegraphics[width=5.5cm,height=5.0cm]{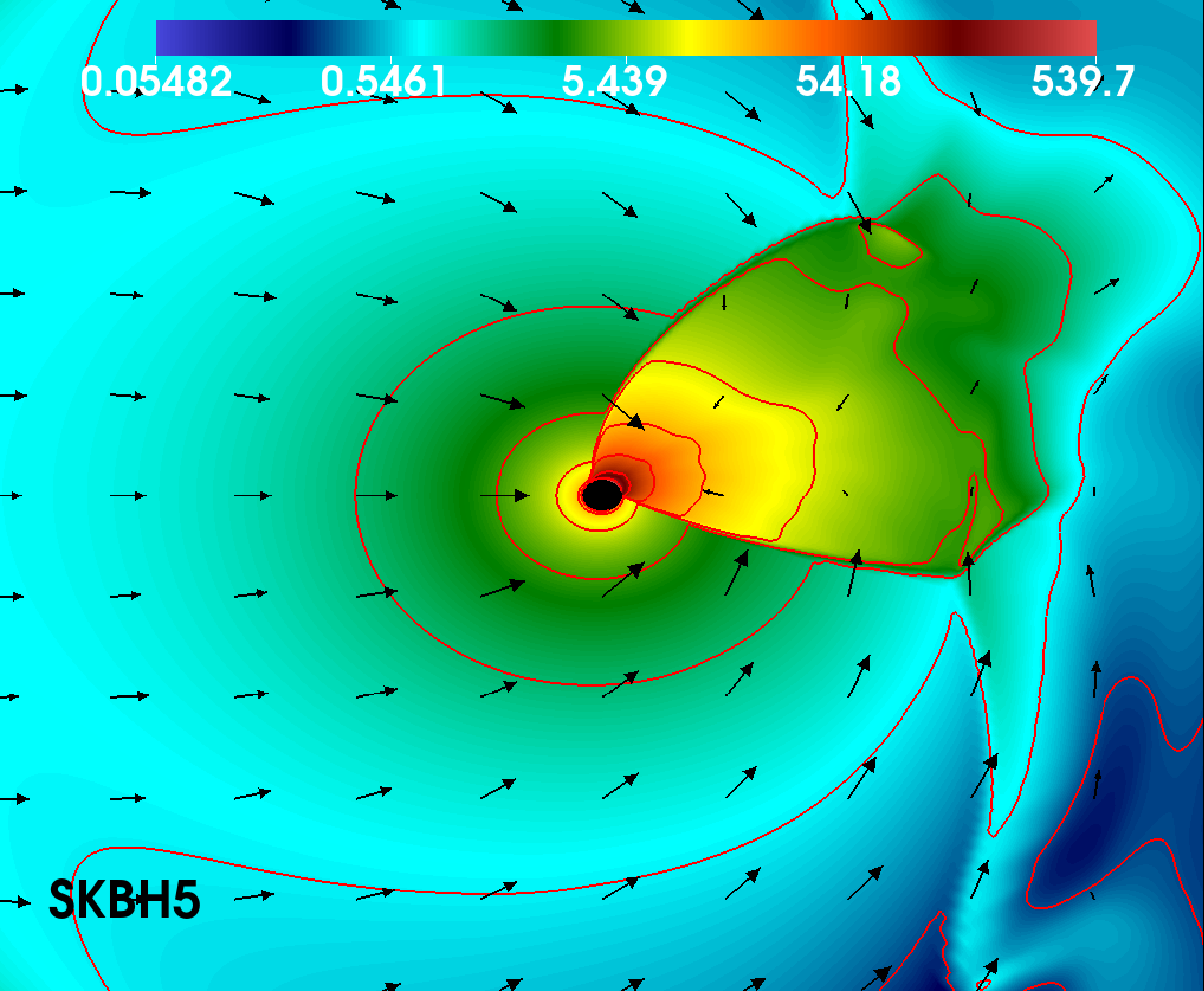}
\includegraphics[width=5.5cm,height=5.0cm]{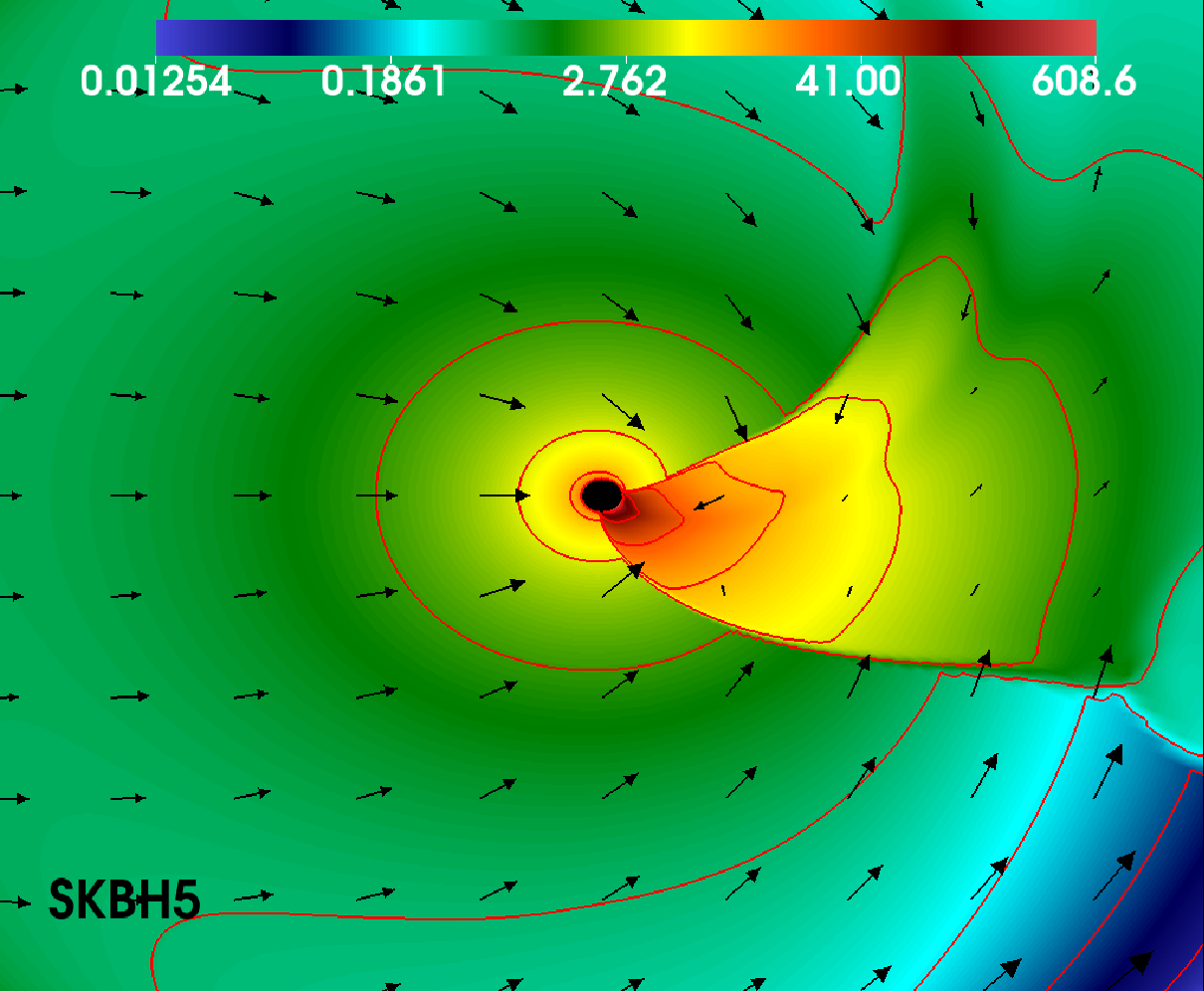}
\includegraphics[width=5.5cm,height=5.0cm]{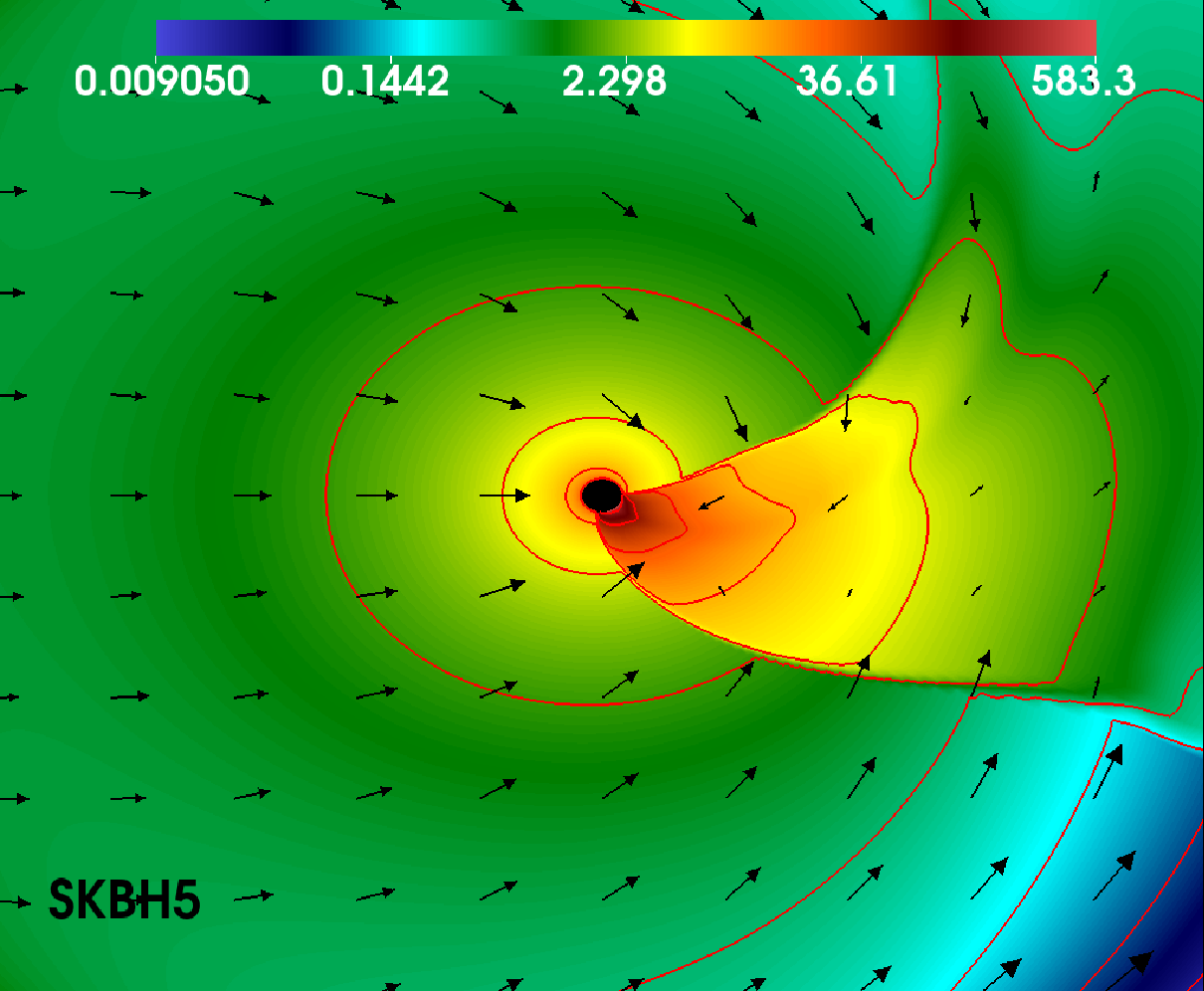}
\caption{Same as Fig.\ref{color_SKBH1}, but for the SKBH5 model given in Table \ref{tab:SKBH_parameters}. In this case, the disformal combination is $C_0-2D_0X_0=1.480$, which is significantly larger than the classical Kerr value. This strong positive deviation from the Kerr solution modifies the downstream shock morphology, producing a warped, shifted, and time-dependent shock-cone/spiral structure around the black hole.}\label{color_SKBH5}
\end{figure*}

In the SKBH5 model shown in Fig.\ref{color_SKBH5}, the accretion morphology is presented for a case that deviates more strongly from both the classical Kerr model and the weakly deviating SKBH1 and SKBH2 models. In this case, unlike the SKBH4 model, the deviation occurs in the positive direction rather than in the negative direction. In this model, the disformal parameters are $C_0=1.50$, $X_0=0.02$, and $D_0=0.50$, giving $C_0-2D_0X_0=1.480$, and the accretion behavior around the slowly rotating disformal black hole is shown. As seen in Fig.\ref{color_SKBH5}, this large positive deformation prevents the formation of a simple, stationary, and well-collimated shock cone in the downstream region. Instead, the shocked matter produces a strongly warped and time-dependent morphological structure, in which its position, opening angle, and orientation change significantly with time. In the snapshots corresponding to the early stages of the simulation, the post-shock region is still observed to form in the downstream region. This shows that the BHL mechanism is active during the early stages of the simulation. However, different from the classical Kerr black-hole model, the wake formed in the downstream region is rapidly distorted, and the high-density region begins to warp around the black hole. After the formation of this mixed shock-cone/spiral morphological structure, a warped shock cone is observed to persist around the black hole for a long time.

A key feature of SKBH5 sen in Fig.\ref{color_SKBH5} is that, when the physical mechanism around the black hole is examined in general, except for the transition time, the shock cone is not completely destroyed, but it is displaced and warped at a different angular position. This situation is clearly seen especially in the last three snapshots at the bottom of Fig.\ref{color_SKBH5}. As seen there, the shock cone is still very dense, but it appears at a different location and continues to oscillate almost periodically throughout the evolution (see in Fig. \ref{mass_accc}) . The cone-like structure alternates between a compact downstream configuration and a broader, curved, spiral-like pattern. On the other hand, from the velocity vector plots, it is seen that the matter in the post-shock region is not only transported away, but also rotates around the black hole or is redirected toward the black hole. This behavior shows that the strong positive deviation seen in $C_0-2D_0X_0$ modifies the effective hydrodynamical response of the accreting matter. Accordingly, the resulting QPO-like behavior is also modified. Compared with the result of the SKBH4 model given in Fig.\ref{color_SKBH4}, in the SKBH5 model shown in Fig.\ref{color_SKBH5}, it is clearly observed that the shock cone is warped, shifted around the black hole, and also undergoes a moderate oscillatory motion. Thus, a different physical region emerges in the SKBH5 model. In other words, the spacetime parameters not only deform or broaden the shock cone, but also generate a recurrent, non-axisymmetric motion of the post-shock region.

\subsection{Azimuthal Density Profiles and Time-Dependent Flow Evolution}
\label{Numeric_3}

\begin{figure*}[tbhp]
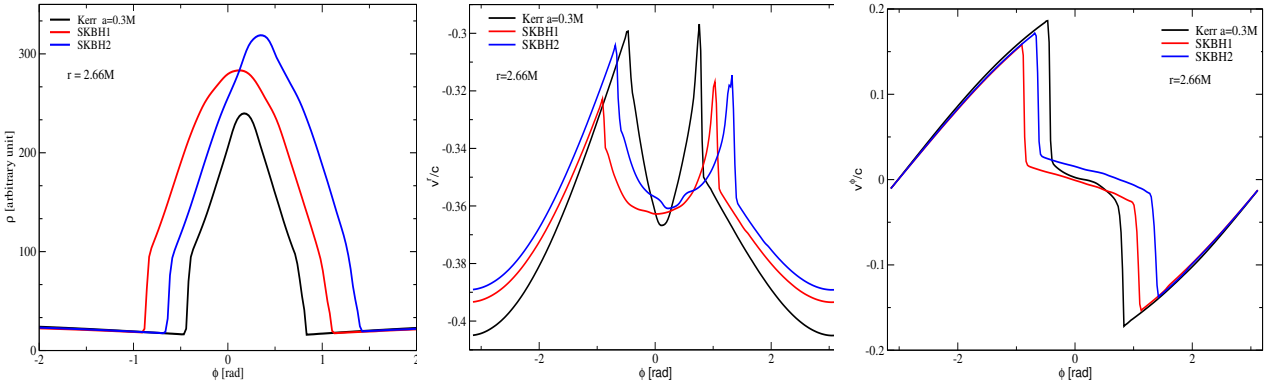

\centering
\includegraphics[width=5.5cm,height=5.0cm]{den_phi_a03_r266.eps}
\includegraphics[width=5.5cm,height=5.0cm]{vr_r_2_66_diff_phi_and_tmaxr.eps}
\includegraphics[width=5.5cm,height=5.0cm]{vphi_r_2_66_diff_phi_and_tmaxr.eps}\\
\caption{One-dimensional azimuthal profiles of the rest-mass density, radial velocity, and angular velocity of the flow morphology formed around the slowly rotating disformal Kerr models SKBH1 and SKBH2 and the Kerr black hole at $r=2.66M$. The profiles show how weak disformal deviations modify the density enhancement, radial inflow transition, and azimuthal motion of the post-shock accreting matter near the black hole.}\label{den_veloc}
\end{figure*}

Here, the one-dimensional profiles of the accretion dynamics around the slowly rotating disformal Kerr black hole are investigated in the azimuthal direction for different models, in the region close to the black-hole horizon, namely in the strong gravitational field. In this way, the parameter-dependent changes of the physical mechanisms formed around the black hole are shown more clearly. In Fig. \ref{den_veloc}, the rest-mass density, radial velocity, and the angular velocity are plotted at $r=2.66M$, in the strong gravitational field, for both the classical Kerr black hole and the rotating disformal Kerr models SKBH1 and SKBH2, with rotation parameter $a=0.3M$, allowing for a direct comparison between them. These one-dimensional cuts clearly reveal how even small changes in the spacetime parameters modify the internal structure of the shock cone and the post-shock flow. In the left panel of Fig. \ref{den_veloc}, it is seen that, in the Kerr$_{03}$ model, the maximum density in the post-shock region of the shock cone is shifted toward the positive $\phi$ region. As discussed in detail in previous studies, this occurs as a result of the warping of the spacetime. On the other hand, compared with the Kerr case, the disformal models produce strong enhancements in the density. The SKBH2 model produces the largest-amplitude peak. This indicates that even weak deviations from the Kerr background can increase the compression of matter inside the shock region. At the same time, the effect of the parameter $C_0-2D_0X_0$ in the disformal spacetime, given in Table \ref{tab:SKBH_parameters}, is clearly seen when compared with Kerr. When $C_0-2D_0X_0$ is smaller than the Kerr value, it produces an effect opposite to the rotation direction of the black hole, while in the model where $C_0-2D_0X_0$ is larger than the Kerr value, it produces an effect in the same direction as Kerr. Therefore, the shift of the maximum amplitude is clearly observed. However, in both disformal models, the opening angle of the resulting shock cone is larger than that of the classical Kerr solution.
The middle panel of Fig. \ref{den_veloc} shows the radial velocity variations for these models. As clearly seen, in the disformal models the radial velocities are shifted and the velocity transitions become broader. This shows that the infalling matter behaves differently after passing through the shock-front region. In the right panel of Fig. \ref{den_veloc}, sharp jumps and displaced discontinuities are observed in the angular velocity, as expected in a standard shock-cone structure. In the disformal models, changes in these features are observed depending on the opening angle of the resulting shock cone. This indicates that the asymmetric azimuthal motion becomes more extended. Thus, the one-dimensional profiles in Fig. \ref{den_veloc} show that not only the morphology of the shock cone changes, but also the compression of the matter trapped inside the cone, the radial inflow structure, and the angular distributions undergo significant changes in the near-black-hole region.

\begin{figure*}[tbhp]
\centering
\includegraphics[width=8.0cm,height=6.0cm]{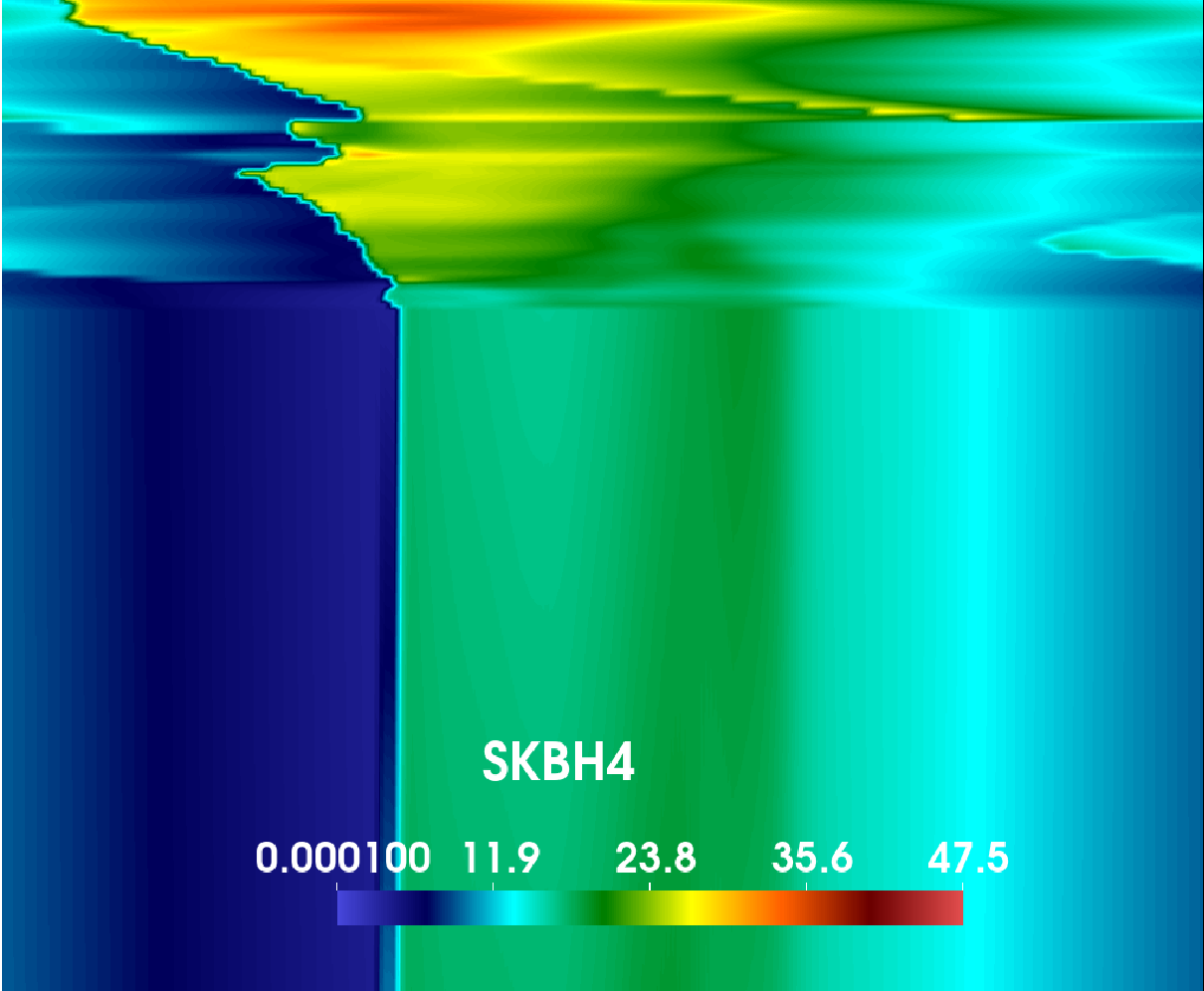}
\includegraphics[width=8.0cm,height=6.0cm]{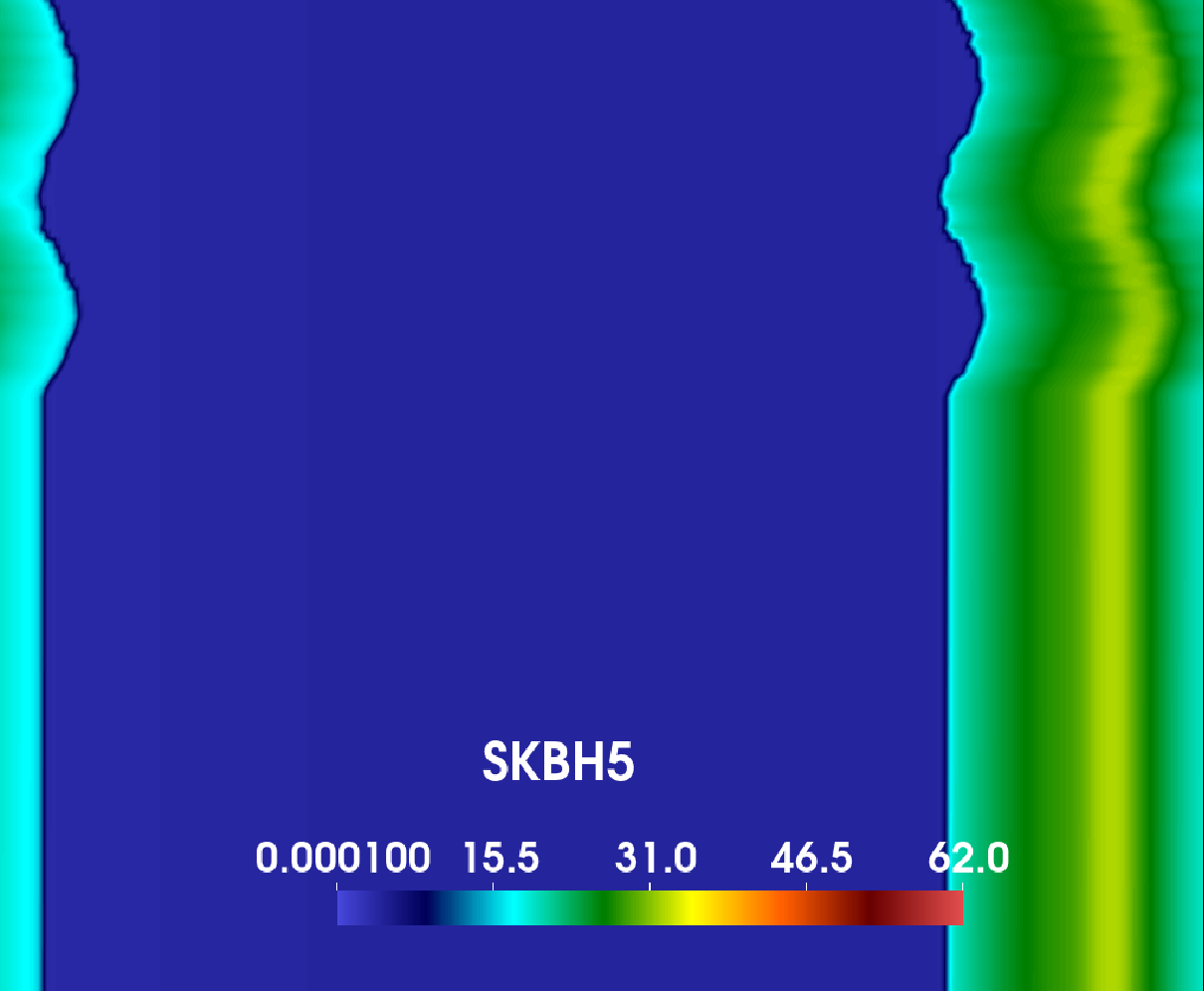}
\caption{Time-dependent azimuthal variation of the rest-mass density for the SKBH4 and SKBH5 models, shown from the stage after the main nonlinear accretion structure has formed until the final time of the simulation. The panels show how the post-shock region evolves with time. While a mixed shock-cone/spiral morphology appears in SKBH4, a more coherent, warped, quasi-periodically oscillating shock structure is observed in the SKBH5 model.}\label{color_t_phi_2}
\end{figure*}

Fig.\ref{color_t_phi_2} presents the time-dependent azimuthal density evolution for the SKBH4 and SKBH5 models after the accretion flow has already developed its main nonlinear structure. Therefore, these panels can be considered as a complementary diagnostic for the snapshots shown in the lower rows of Figs.\ref{color_SKBH4} and \ref{color_SKBH5}. In other words, for these models, the time evolution of the post-shock region is revealed from the later stage of the accretion morphology formed around the black hole until the final stage of the simulation. For the SKBH4 model, the density map shows a broader, irregular, and strongly destroyed structure. This indicates that the strong negative deviation from the classical Kerr solution leads to the formation of a configuration that is very different from the classical shock oscillation. At the same time, it confirms the formation of a mixed shock-cone and spiral morphology. The density interface is not vertical or stationary. Instead, it bends and drifts with time. This shows that the shocked matter around the black hole is continuously redistributed, as seen in Fig.\ref{color_SKBH4}. In contrast, the SKBH5 model shows a more coherent behaviour, but unlike the classical Kerr solution, it exhibits a time-dependent and axisymmetric-like distribution around $\phi=0\,\mathrm{rad}$. This is consistent with the behavior of the warped shock cone shown in Fig.\ref{color_SKBH5} . At the same time, after a certain time, a QPO is observed in the right panel of Fig.\ref{color_t_phi_2}. This shows that large negative and positive deviations generate different hydrodynamical regions. The time-dependent azimuthal variations shown in Fig.\ref{color_t_phi_2} are physically important because these behaviors directly control the variation of the mass accretion rate and provide details about the hydrodynamical origin of the QPO-like variations discussed later.

\section{Timing Signatures of BHL Accretion: Accretion-Rate Variability and QPO Frequencies}
\label{Numeric_4}
In this section, rather than focusing on the spatial behavior of the flow morphology, we reveal how it evolves with time and how it can be connected to observability. The azimuthal profiles and two-dimensional density maps discussed in Section \ref{Numeric_1} show that the disformal parameters strongly modify the shock cone around the black hole, produce spiral-like overdense regions, and generate non-axisymmetric time-dependent motions. The spacetime parameters not only change the morphological properties of the dynamical structures, but also cause variations in the amount of matter falling through the inner boundary toward the black hole. Thus, the direct signatures of these time-dependent variations are seen in the mass accretion rate. For this reason, the temporal behavior of the accretion rate provides a powerful diagnostic for connecting the hydrodynamical response of the BHL flow to observable quasi-periodic variability.

The aim of this section is to quantify how different disformal Kerr models transfer nonlinear shock dynamics into timing signals. For this purpose, we first reveal the variations in the mass accretion rate for the Kerr, SKBH1, SKBH2, SKBH4, and SKBH5 models, and then compute the PSD analyses from these mass accretion rates. By fitting the dominant PSD features with Lorentzian components, we obtain the characteristic QPO-like frequencies for each spacetime configuration. Thus, the oscillating shock cone, warped post-shock structure, and spiral-density modulations are transferred into a measurable frequency domain. In this way, we reveal what kind of measurable effects the spacetime parameters used here produce in accreting black-hole systems.

\subsection{Mass Accretion-Rate Variability Across Disformal Kerr Models}
\label{Numeric_5}

\begin{figure*}[tbhp]
\centering
\includegraphics[width=15.5cm,height=14.0cm]{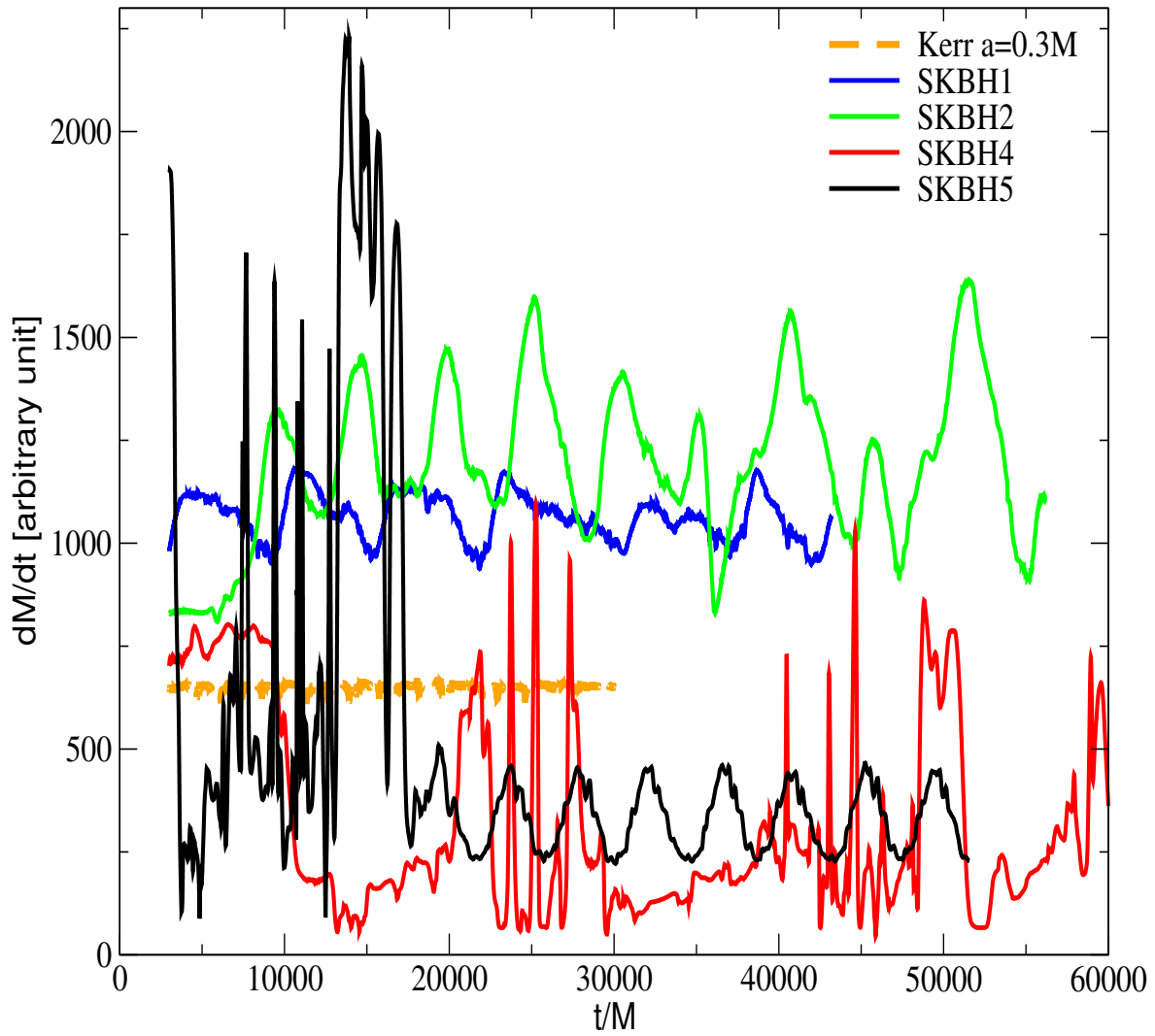}
\caption{Time evolution of the mass accretion rates calculated at the inner boundary around the Kerr black hole together with the slowly rotating disformal Kerr models SKBH1, SKBH2, SKBH4, and SKBH5 for $a=0.3M$. While an almost steady accretion occurs in the Kerr case plotted as the reference model, the disformal models generally show significantly enhanced amplitudes and model-dependent behaviors. These differences show how the disformal spacetime parameters imprint the shock-cone, spiral, and warped post-shock dynamics directly onto the accretion-rate signal.}\label{mass_accc}
\end{figure*}

In Fig. \ref{mass_accc}, the time-dependent mass accretion rates are given at the inner boundary of the computational domain, namely in the region closest to the black-hole horizon, for the slowly rotating disformal Kerr models SKBH1, SKBH2, SKBH4, and SKBH5, together with the Kerr model. For the Kerr model, the accretion rate remains almost steady with small oscillations compared with the other models. It produces weak oscillations around an almost constant value. This behavior is consistent with the standard BHL picture. Compared with the disformal cases, the Kerr case is used as the reference level. Thus, the variations in the other disformal models and the effects produced by the spacetime can be clearly revealed.

In the SKBH1 model shown in Fig. \ref{mass_accc}, the disformal deviation is weak and can be classified as negative with respect to the Kerr reference case. The accretion rate is larger than that of the Kerr case and shows a moderate oscillatory behavior. This is consistent with Fig.\ref{color_SKBH1}, where the shock cone still exists, has a broader structure, and oscillates in the azimuthal direction. Thus, the mass accretion-rate curve shows that even small changes in the Kerr background can cause a significant increase in the amount of matter captured by the black hole. At the same time, it produces a time-dependent oscillatory modulation. However, the variability is still relatively smooth compared with the more strongly deformed models.

In the SKBH2 model shown in Fig. Fig. \ref{mass_accc}, the mass accretion rate exhibits stronger variability and has a larger amplitude than those of the Kerr and SKBH1 models. This behavior is consistent with the results shown in Figs.\ref{color_SKBH2} and \ref{den_veloc}. This is because the SKBH2 model shows stronger density enhancement, a wider shock opening angle, and a clear deformation of the shock cone in the downstream region. The large-scale oscillations that vary with time repeat regularly. This shows that the shock cone does not only fluctuate irregularly, but also produces persistent quasi-periodic modulations in the accretion flow. For this reason, the SKBH2 model, which has a small positive deviation with respect to the Kerr model, appears as one of the clearest models.

In the SKBH4 model shown in Fig. \ref{mass_accc}, the accretion rate is strongly irregular and intermittent. In this curve, sharp spikes and strong bursts are observed, especially at later times of the simulation. This is directly consistent with the accretion morphology given in Fig.\ref{color_SKBH4} and the azimuthal density variation shown in Fig.\ref{color_t_phi_2}. In the SKBH4 model, the classical shock cone is strongly distorted and transformed into a mixed shock-cone/spiral structure. Since the shocked matter is rearranged around the black hole in an unstable and non-axisymmetric way, the amount of matter falling toward the black hole changes abruptly. Thus, the strong negative deviation from the classical Kerr solution reflects the burst-like accretion-rate behavior of SKBH4 as a strongly nonlinear and unstable hydrodynamical response.

In the SKBH5 model shown in Fig. \ref{mass_accc}, the mass accretion rate exhibits a very violent time-dependent variation, especially during the early stages of the simulation. The amplitude of the variations in this phase is very large. However, at later times, it continues to oscillate with a lower amplitude. This behavior is consistent with the two-dimensional flow morphology shown in  Fig.\ref{color_SKBH5}. This behavior in the mass accretion rate again indicates that there is a strong transition phase during the early stage of the simulation. Later, it confirms that an axisymmetric-like shock cone around $\phi=0\,\mathrm{rad}$ and, at the same time, a warped oscillating shock cone around the black hole are formed in the downstream region. The late-time behavior of the mass accretion rate also confirms the coherent quasi-periodic azimuthal density pattern seen in Fig.\ref{color_t_phi_2}. Thus, the SKBH5 model shows a different temporal regime from the SKBH4 model. Instead of producing only irregular burst-like behavior, the strong positive disformal deviation generates a warped shock structure that can repeatedly modulate the accretion rate.

\subsection{PSD-Based Identification of QPO Signatures}
\label{Numeric_6}
\begin{figure*}[tbhp]
\centering
\includegraphics[width=8.0cm,height=7.0cm]{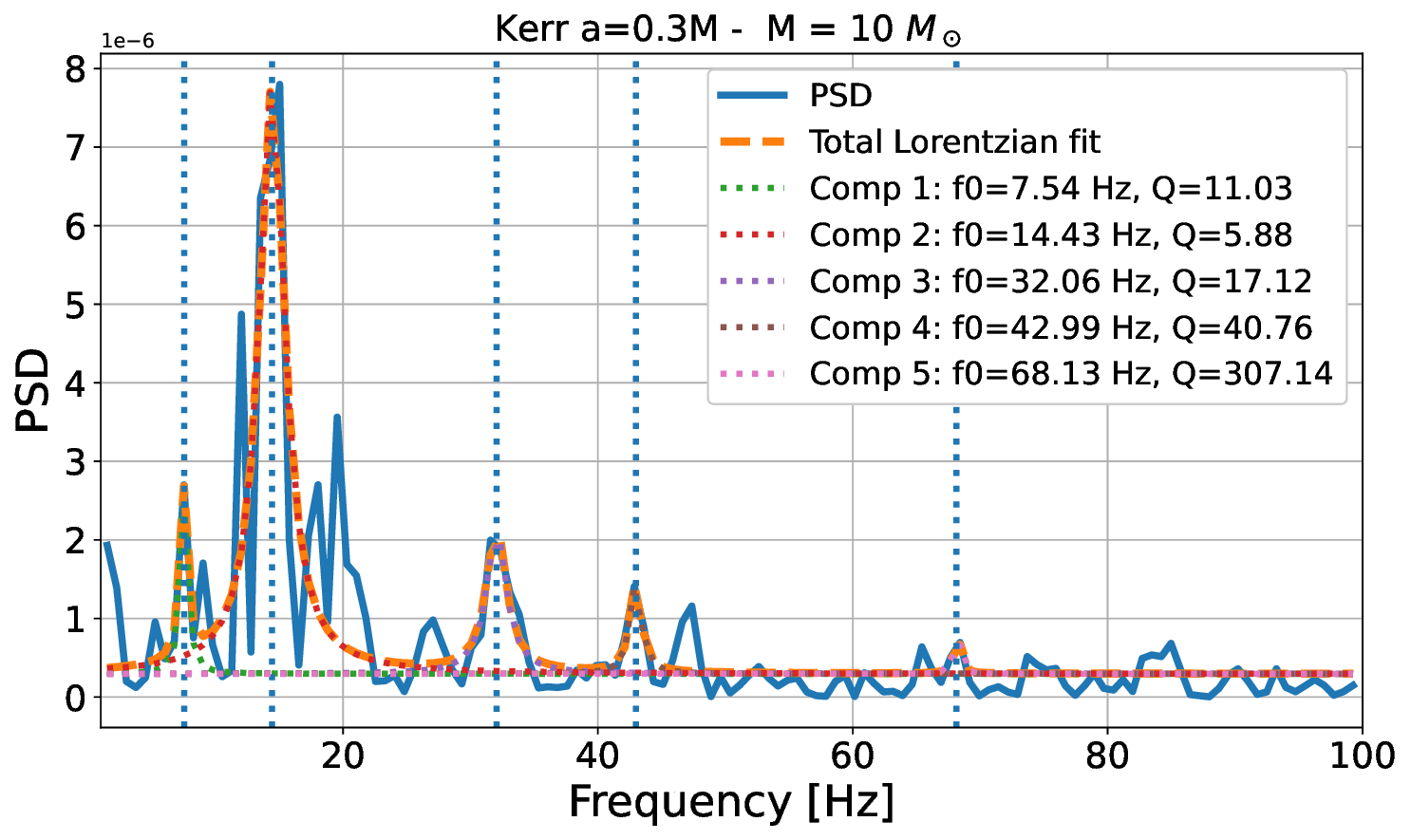}
\includegraphics[width=8.0cm,height=7.0cm]{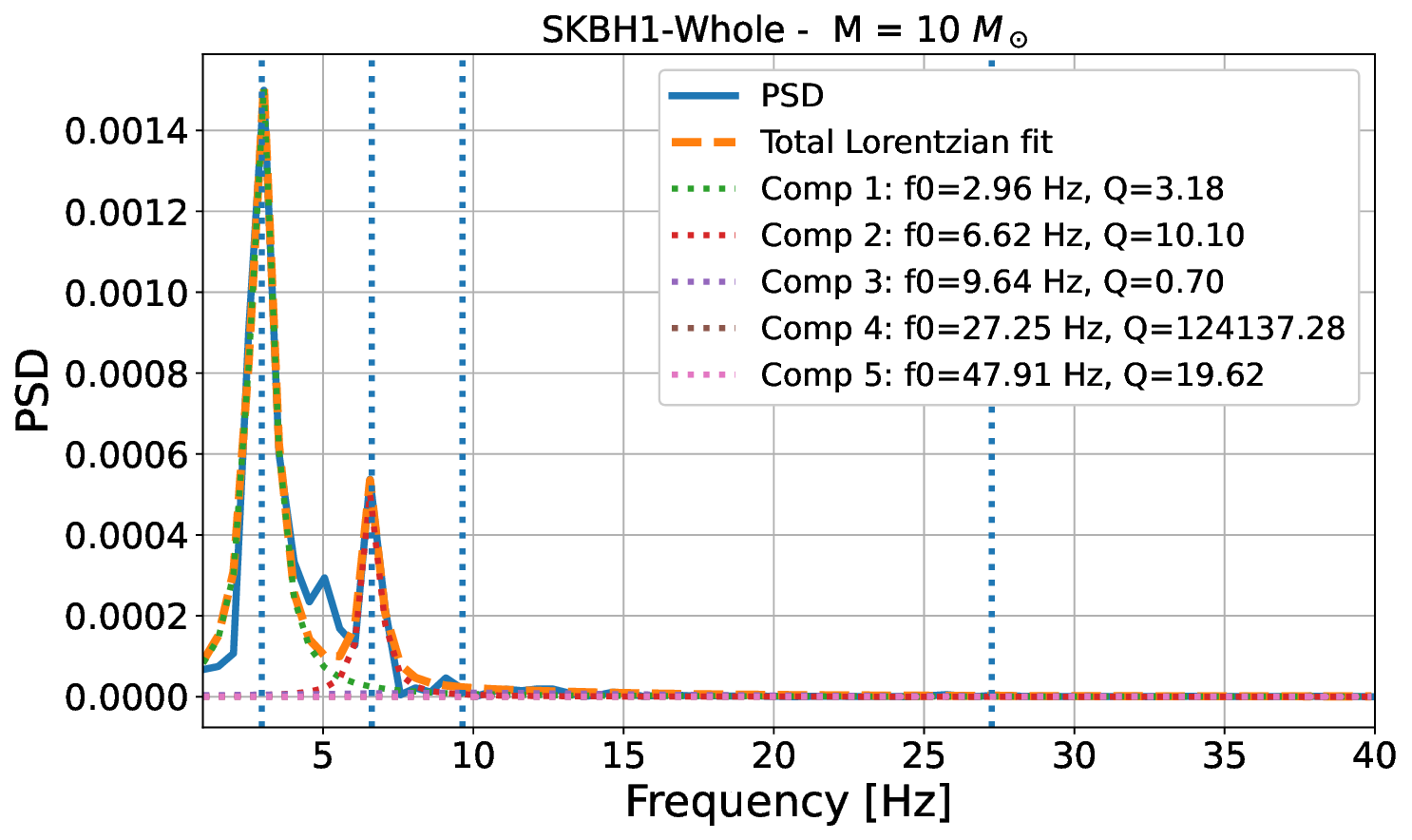}\\
\includegraphics[width=8.0cm,height=7.0cm]{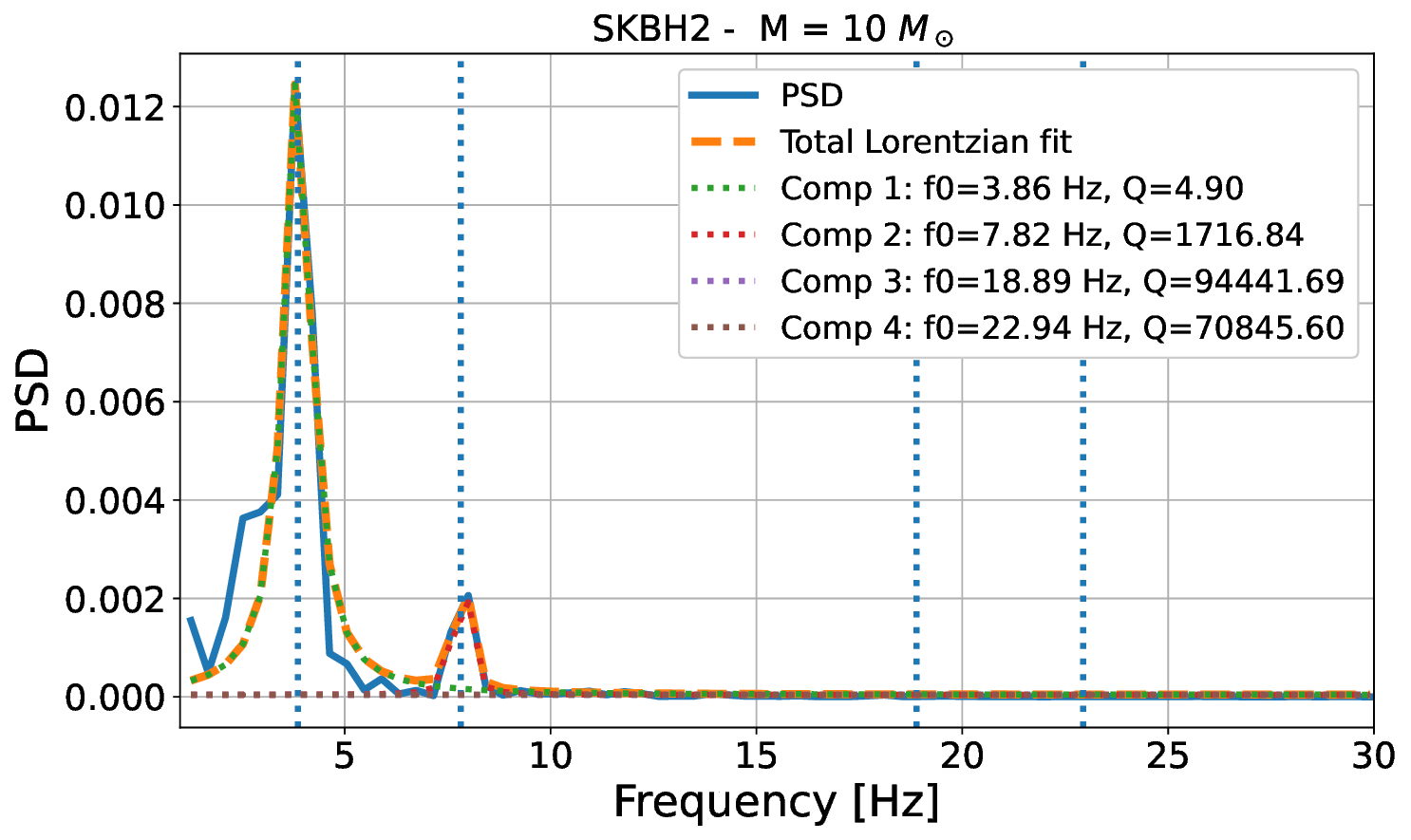}
\includegraphics[width=8.0cm,height=7.0cm]{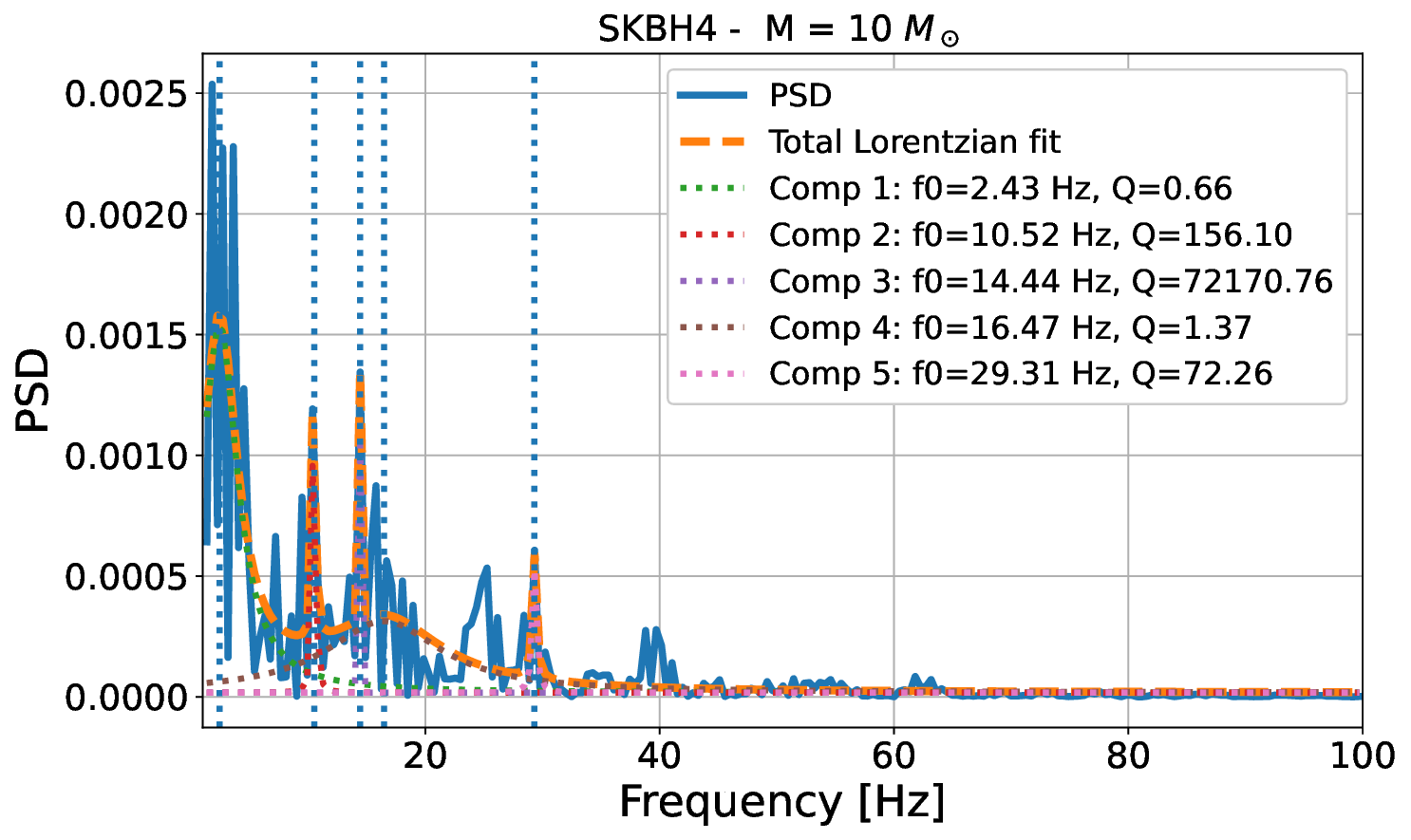}
\caption{PSD spectra and multi-component Lorentzian fits calculated from the mass accretion rate at $r=2.3M$ are shown for both the Kerr model and the slowly rotating disformal models SKBH1, SKBH2, and SKBH4. The blue curve represents the numerically calculated PSD, while the Lorentzian components show the dominant QPO-like peaks and their coherence properties. Compared with the classical Kerr solution, the disformal models shift the timing power and modify the peak structures. The physical mechanisms formed around the slowly rotating disformal Kerr black hole generate QPO signatures associated with the shock cone, warped post-shock flow, and mixed shock-cone/spiral morphology.}\label{PSD_1_2_4}
\end{figure*}

In Fig.\ref{PSD_1_2_4}, the PSD spectra and multi-component Lorentzian fits are shown for the Kerr model with rotation parameter $a=0.3M$ and for the disformal models SKBH1, SKBH2, and SKBH4. Fig.\ref{PSD_1_2_4} converts the time-domain variations shown in Fig. \ref{mass_accc} into the frequency domain, and thus we reveal the characteristic QPO-like frequencies numerically produced by the shock cone, warped post-shock flow, and spiral-density structures. In the profiles, the blue curve represents the numerically calculated PSD, while the Lorentzian components decompose the dominant peaks, determine their central frequencies, and reveal their coherence properties by calculating the factor $Q$.

In the upper-left panel of  Fig.\ref{PSD_1_2_4}, the PSD analysis and Lorentzian components are shown for the Kerr reference model. It is seen that the PSD produces several Lorentzian peaks, and these peaks occur approximately at $7.54\,\mathrm{Hz}$, $14.43\,\mathrm{Hz}$, $32.06\,\mathrm{Hz}$, $42.99\,\mathrm{Hz}$, and $68.13\,\mathrm{Hz}$. While the strongest low-frequency structure appears at $14.43\,\mathrm{Hz}$, the peaks at $42.99\,\mathrm{Hz}$ and $68.13\,\mathrm{Hz}$ have the largest $Q$ values. This shows that these peaks with larger $Q$ values are narrower and more coherent oscillatory components. These results are consistent with the behavior of the mass accretion rate discussed in Fig. \ref{mass_accc}, which is almost steady but weakly oscillatory. In the Kerr model, the resulting shock cone is relatively stable. Therefore, the Kerr model does not produce a single overwhelmingly dominant low-frequency modulation. Instead, the radial and azimuthal modes trapped inside the shock cone are gradually excited during the evolution, producing multiple characteristic frequencies through the fundamental modes and their nonlinear coupling. This situation has been discussed in detail in our previous studies \cite{Donmez2024JCAP, Donmez2025JHEAp}. As a result of these physical processes, several relatively weak peaks distributed over a broad frequency range are formed.

In the upper-right panel of Fig.\ref{PSD_1_2_4}, for the SKBH1 model, although the deviation from the Kerr solution is small, significant changes are observed in the PSD. The dominant Lorentzian peaks are observed at $2.96\,\mathrm{Hz}$, $6.62\,\mathrm{Hz}$, $9.64\,\mathrm{Hz}$, $27.25\,\mathrm{Hz}$, and $47.91\,\mathrm{Hz}$, while it is clearly seen that the strongest powers occur especially at the low frequencies $2.96\,\mathrm{Hz}$ and $6.62\,\mathrm{Hz}$. The quality factor of the $6.62\,\mathrm{Hz}$ component is larger than that of the $2.96\,\mathrm{Hz}$ component. This shows that $6.62\,\mathrm{Hz}$ is a more coherent oscillation mode. Although the peak at $27.25\,\mathrm{Hz}$ is seen to be very narrow and coherent, its power is much lower compared with the $2.96\,\mathrm{Hz}$ and $6.62\,\mathrm{Hz}$ peaks, which significantly reduces its observability. Compared with the Kerr solution, it is observed that the timing power in the SKBH1 model shifts toward lower frequencies. This is consistent with the flow morphology shown in Fig.\ref{color_SKBH1}. As discussed in Fig.\ref{color_SKBH1}, the shock-cone structure is still preserved, but the cone becomes broader and its azimuthal oscillation is strengthened. Thus, the PSD analysis for SKBH1 shows that a weak negative disformal deviation can transform a relatively steady Kerr-like shock cone into a more slowly oscillating post-shock structure.

In the lower-left panel of Fig.\ref{PSD_1_2_4}, for the SKBH2 model, the PSD is organized in a sharper form. The main Lorentzian peaks occur at $3.86\,\mathrm{Hz}$, $7.82\,\mathrm{Hz}$, $18.89\,\mathrm{Hz}$, and $22.94\,\mathrm{Hz}$. While the strongest feature is formed at the low frequency of $3.86\,\mathrm{Hz}$, the higher-frequency peaks with lower power, especially at $18.89\,\mathrm{Hz}$ and $22.94\,\mathrm{Hz}$, have very large $Q$ values. Although these peaks are very narrow and coherent, their observability is much lower than that of the other two peaks. On the other hand, the peak at $7.82\,\mathrm{Hz}$ also has a high $Q$ value and an observable level of power. Therefore, due to its strong coherence and very narrow structure, it appears as a suitable component for observations. The PSD behavior found in this model is consistent with the persistent and strong quasi-periodic modulation shown by the mass accretion-rate curve of SKBH2 discussed in Fig. \ref{mass_accc}. At the same time, it is also consistent with the density and velocity diagnostics given in Figs. \ref{color_SKBH2} and \ref{den_veloc}. Thus, among the weak-deviation models, SKBH2 provides the clearest example of how a small positive deviation from Kerr can produce a stable and coherent QPO-like frequency pattern.

Finally, in the lower-right panel of Fig.\ref{PSD_1_2_4}, the PSD analysis of the SKBH4 model is spread over a broader frequency range and is more complex. The Lorentzian components occur approximately at $2.43\,\mathrm{Hz}$, $10.52\,\mathrm{Hz}$, $14.44\,\mathrm{Hz}$, $16.47\,\mathrm{Hz}$, and $29.31\,\mathrm{Hz}$. Unlike the SKBH1 and SKBH2 models, in the PSD analysis of the SKBH4 model, several strong peaks distributed over a broad frequency range are observed. The components at $2.43\,\mathrm{Hz}$ and $16.47\,\mathrm{Hz}$ have relatively low $Q$ values. This shows that these peaks are broader and almost non-coherent. On the other hand, the components at $10.52\,\mathrm{Hz}$, $14.44\,\mathrm{Hz}$, and $29.31\,\mathrm{Hz}$ indicate more coherent structures, which are embedded inside an overall turbulent signal. These PSD and Lorentzian-fit results are fully consistent with the morphological behavior of the SKBH4 model observed in Figs.\ref{color_SKBH4} and \ref{color_t_phi_2}. As known in section \ref{Numeric_1}, in this model, the classical shock cone is completely destroyed and replaced by a mixed shock-cone/spiral structure. Therefore, the SKBH4 PSD reflects a combination of irregular burst-like accretion and localized coherent oscillatory modes.

\begin{figure*}[tbhp]
\centering
\includegraphics[width=8.0cm,height=7.0cm]{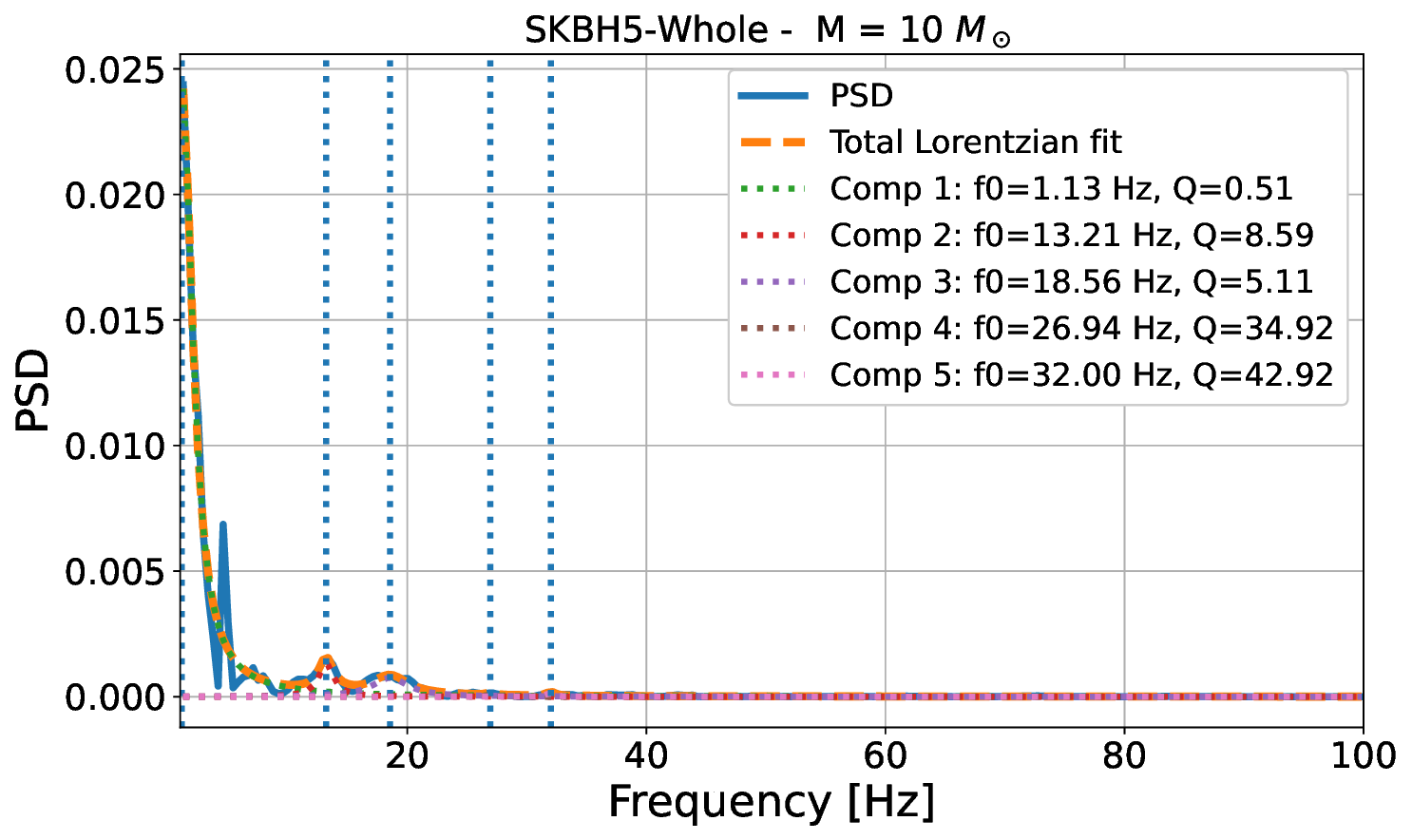}
\includegraphics[width=8.0cm,height=7.0cm]{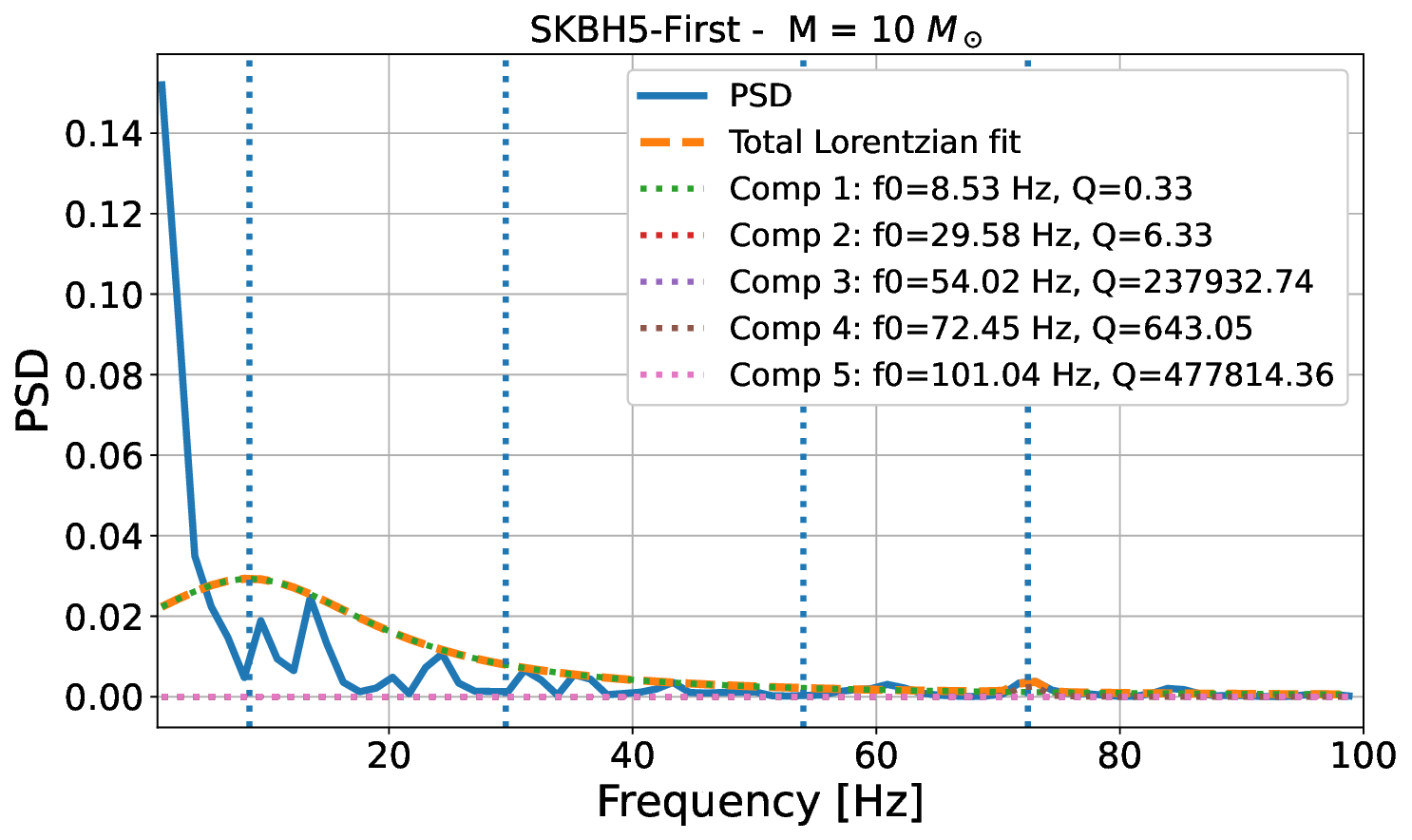}
\includegraphics[width=8.0cm,height=7.0cm]{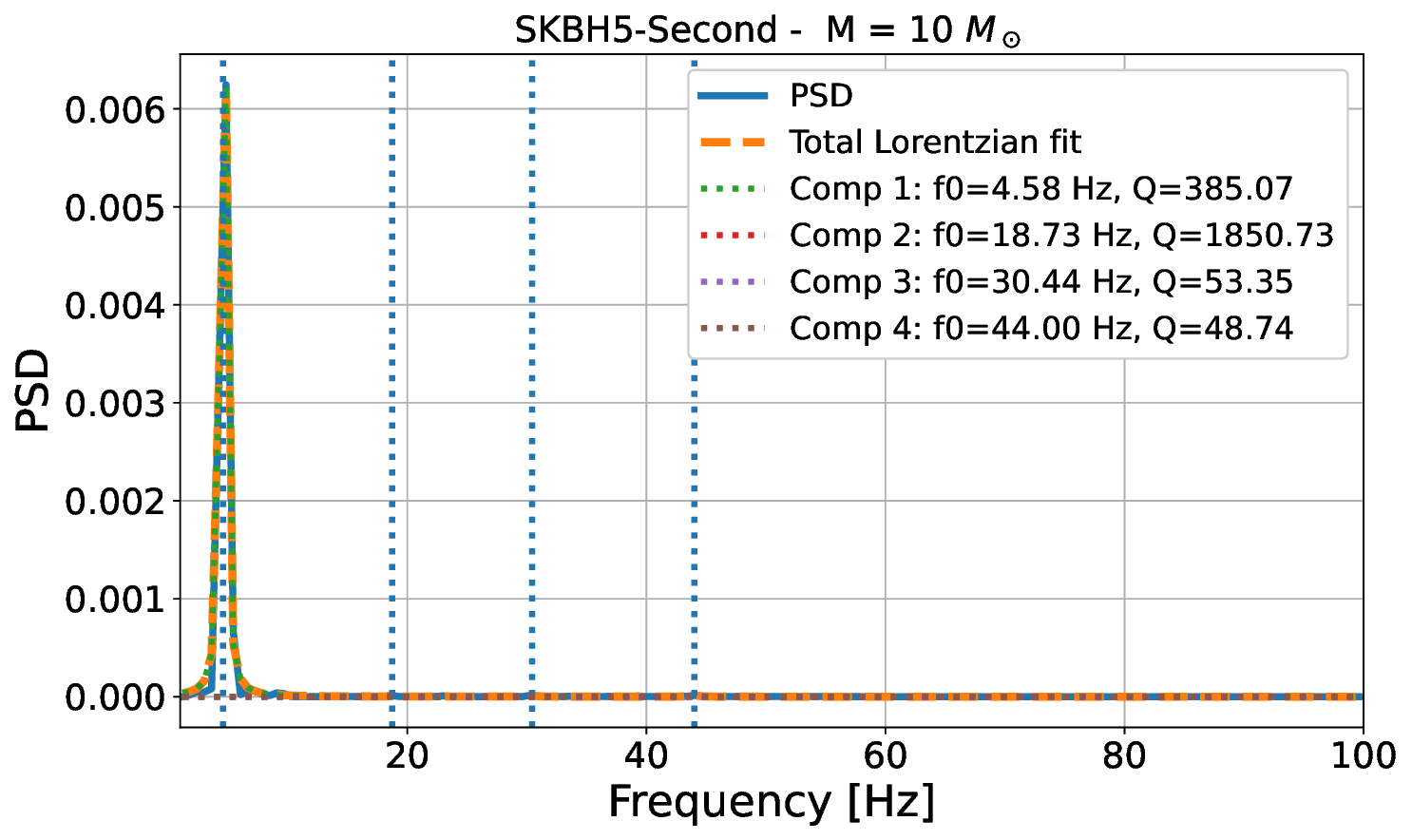}
\caption{It is the same as Fig.\ref{PSD_1_2_4}, but this time for the SKBH5 model. Here, three different PSD analyses are presented because two different physical states appear with time in the SKBH5 model. The upper-left panel shows the PSD obtained from the mass accretion rate calculated over the whole simulation, while the upper-right panel shows the PSD analysis of the initial stage of the simulation, namely the transition region where strong oscillations occur. The lower panel shows the PSD obtained from the later phase of the simulation, after the warped shock cone has formed. The Lorentzian components show the QPO-like peaks produced by these different hydrodynamical phases.
}\label{PSD_SKBH5}
\end{figure*}

In Fig. \ref{PSD_SKBH5}, we perform the PSD analysis and Lorentzian fitting only for the SKBH5 model. Since two different hydrodynamical states appear during the simulation in this model, we discuss the possible spectral states observed in each hydrodynamical phase. In the upper-left panel of Fig. \ref{PSD_SKBH5}, the PSD is calculated from the mass accretion rate over the whole simulation. Therefore, it contains a mixed PSD analysis of the initial transition phase and the later warped-shock oscillation state. As a result, the Lorentzian components are observed around $1.13\,\mathrm{Hz}$, $13.21\,\mathrm{Hz}$, $18.56\,\mathrm{Hz}$, $26.94\,\mathrm{Hz}$, and $32.00\,\mathrm{Hz}$. However, since the low-frequency component has a small $Q$ value, it is seen that the signal produces broad and non-coherent variability due to the dominance of the transition phase. In the upper-right panel of Fig. \ref{PSD_SKBH5}, the PSD analysis of the initial transition part up to $t=20000M$ in the mass accretion rate shown in Fig. \ref{mass_accc} is presented, corresponding to the region where large-amplitude accretion oscillations occur. In this interval, the PSD is dominated by a broad low-frequency structure located approximately at $8.53\,\mathrm{Hz}$. Since its $Q$ value is very small, it does not produce a coherent oscillation. At the same time, peaks around $29.58\,\mathrm{Hz}$, $54.02\,\mathrm{Hz}$, $72.45\,\mathrm{Hz}$, and $101.04\,\mathrm{Hz}$ are also observed in the PSD analysis of the same interval. This shows that the early phase of the SKBH5 model is controlled by the violent and nonlinear restructuring of the accretion dynamics. Therefore, the resulting QPO modes are not clear and single. This behavior is consistent with Fig.\ref{color_SKBH5}. In contrast, the lower panel of Fig. \ref{PSD_SKBH5} represents the later phase of the simulation, in which the system forms a warped shock cone and oscillates around the black hole with a clear period. In this region, the PSD peaks are more coherent and narrower. The resulting Lorentzian components occur around $4.58\,\mathrm{Hz}$, $18.73\,\mathrm{Hz}$, $30.44\,\mathrm{Hz}$, and $44.00\,\mathrm{Hz}$. In this case, the peak at $4.58\,\mathrm{Hz}$ has both a very strong amplitude and a strongly coherent, narrow structure. Therefore, it appears as the strongest observable QPO-like structure. The strongest physical message is that SKBH5 produces two distinct timing states: an early transient state with violent, broad-band accretion variability and a later quasi-periodic state associated with the recurrent motion of the warped shock cone. Thus, Fig. \ref{PSD_SKBH5} shows that the strong positive disformal deviation in the SKBH5 model does not only shift the QPO frequencies, but also changes the entire temporal character of the accretion flow. This is produced by two different physical mechanisms: a burst-like transition-phase dynamics and a more coherent late-time warped shock-cone oscillation phase.

\begin{table*}[ht]
\centering
\small
\setlength{\tabcolsep}{4pt}
\caption{Comparison between observed QPO frequencies, observed black-hole masses, numerical frequencies, and the numerically expected black-hole masses obtained from the inverse mass-scaling relation. The numerical frequencies were calculated for a reference black-hole mass of $10M_{\odot}$.}
\label{qpo_observational_counterparts}
\begin{tabular}{l c c c c c}
\hline
Observed system & 
Observed QPO frequency & 
Observed mass & 
Numerical model & 
Numerical frequency used & 
Numerically expected mass \\
\hline

GRS 1915+105 & 
$41$ and $67~{\rm Hz}$ & 
$12.4^{+2.0}_{-1.8}M_{\odot}$ & 
Kerr & 
$42.99$ and $68.13~{\rm Hz}$ & 
$10.49$ and $10.17M_{\odot}$ \\

GRO J1655--40 & 
$300$ and $450~{\rm Hz}$ & 
$7.02 \pm 0.22M_{\odot}$ & 
Kerr & 
$42.99$ and $68.13~{\rm Hz}$ & 
$1.43$ and $1.51M_{\odot}$ \\

XTE J1550--564 & 
$184$ and $276~{\rm Hz}$ & 
$\sim 9.41M_{\odot}$ & 
Kerr & 
$42.99$ and $68.13~{\rm Hz}$ & 
$2.34$ and $2.47M_{\odot}$ \\

Any BH LFQPOs & 
$0.1$--$30~{\rm Hz}$ & 
Source class & 
SKBH1--SKBH5 & 
$2$--$30~{\rm Hz}$ & 
$\sim 3.1$--$3000M_{\odot}$ \\

M82 X--1 \cite{Mondal:2022hle} & 
$3.3$ and $5.1~{\rm Hz}$ & 
$\sim 156$--$380M_{\odot}$ & 
Kerr & 
$42.99$ and $68.13~{\rm Hz}$ & 
$130.27$ and $133.59M_{\odot}$ \\

NGC 5408 X--1 \cite{2009ApJ...703.1386S} & 
$20~{\rm mHz}$ or $15~{\rm mHz}$ & 
$\sim 1000$--$9000M_{\odot}$ & 
SKBH2/SKBH5 & 
$3.86$ / $4.58~{\rm Hz}$ & 
$1930$ and $3053M_{\odot}$ \\

RE J1034+396 \cite{2010MNRAS.401..507B} & 
$0.00027~{\rm Hz}$ & 
$\sim (1$--$4)\times 10^{6}M_{\odot}$ & 
SKBH5/Kerr & 
$4.58$ or $68.13~{\rm Hz}$ & 
$1.70\times 10^{5}$ or $2.52\times 10^{6}M_{\odot}$ \\

\hline
\end{tabular}
\end{table*}

\section{Mass-Scaled QPO Interpretation and Observational Counterparts of Simulated BHL Variability}
\label{Numeric_7}
The QPO-like frequencies obtained from the Lorentzian decomposition provide a direct way to compare the numerically modeled BHL variability with observed black-hole timing signals. Since the numerical frequencies in this work are computed for a black-hole mass of $10M_{\odot}$, the comparison with other systems can be made through the standard inverse-mass scaling of dynamical frequencies. Mass scaling is necessary because the characteristic dynamical frequencies near the black hole scale approximately as one over the black-hole mass. Therefore, the frequencies found in this work for a black hole with $M=10M_{\odot}$ can be recalculated for black holes with different masses by using the following relation: 

\bea
f_{\rm obs}=f_{\rm sim}\left(\frac{10M_{\odot}}{M_{\rm BH}}\right).
\label{fobs}
\eea

\noindent 
Similarly, the black-hole mass can be estimated by using the following formula:

\bea
M_{\rm BH}=10M_{\odot}\left(\frac{f_{\rm sim}}{f_{\rm obs}}\right).
\label{MBH}
\eea

\noindent 
Through this scaling, the same hydrodynamical mechanism can be applied to stellar-mass black-hole binaries, ULXs, and AGN.

Thus, the hydrodynamical mechanism that produces frequencies from a few Hz to tens of Hz for a black hole with $M=10M_{\odot}$ produces frequencies at the mHz level when applied to intermediate-mass black-hole sources, while for massive black-hole sources it produces frequencies in the range from $0.0001\,\mathrm{Hz}$ down to $1\,\mu\mathrm{Hz}$. Therefore, the QPO results numerically obtained here can be used not only to compare with Galactic X-ray binary systems, but also to compare the numerical and observational results by applying them to ULXs and AGN QPO candidates. For example, the QPO-like frequency of $4.58\,\mathrm{Hz}$ found for a black hole with $M=10M_{\odot}$ becomes $45.8\,\mu\mathrm{Hz}$ for the same mode around a black hole with mass $M=10^6M_{\odot}$. This corresponds to a period of about $6.1$ hours. If the black-hole mass is $M=10^7M_{\odot}$, the same mode becomes $4.58\,\mu\mathrm{Hz}$, and the corresponding period is $2.5$ days. Thus, if the mass of the observed source is known, the numerical frequencies found in this work can be directly shifted to the expected observed frequencies by using inverse-mass scaling.

When the numerical results are compared with observations seen in Table \ref{qpo_observational_counterparts}, the Kerr reference model produces results that can be more directly compared with HFQPOs observed in Galactic black-hole binaries. In particular, the coherent Kerr peaks at $42.99\,\mathrm{Hz}$ and $68.13\,\mathrm{Hz}$ are close to the well-known QPO frequencies of approximately $41\,\mathrm{Hz}$ and $67\,\mathrm{Hz}$ observed from the source GRS 1915+105. Belloni and collaborators discuss the recurrent approximately $67\,\mathrm{Hz}$ feature and the weaker approximately $41\,\mathrm{Hz}$ detection in averaged RXTE observations, making this source the closest stellar-mass counterpart to the Kerr reference frequencies obtained here \cite{Orh2_1, belloni2012high}. This close relation shows that sources similar to GRS 1915+105 can exhibit high-frequency timing behavior through the standard Kerr-like BHL shock-cone oscillation without requiring any strong disformal deviation.

The same frequency pair numerically found in the Kerr model produces a mode coupling close to the $3:2$ ratio. This is very important because several observed black-hole binary systems show this type of twin high-frequency QPO ratio. For example, the best-known pair observed from the source GRO J1655-40 is $300\,\mathrm{Hz}$ and $450\,\mathrm{Hz}$. On the other hand, simultaneous HFQPOs around $184\,\mathrm{Hz}$ and $276\,\mathrm{Hz}$ have also been observed from the source XTE J1550-564 \cite{2001ApJ...552L..49S, Remillard:2006fc}. These observed frequencies are larger than the numerical frequencies calculated from the Kerr model for a black hole with $M=10M_{\odot}$. Therefore, these observational results and numerical results cannot be directly matched because the frequencies do not agree exactly. However, the similarity in the frequency ratio indicates that the nonlinear coupling of modes in the simulated shock-cone system may produce the same type of harmonic structure that is commonly discussed in the observational literature.

Weak disformal models are naturally connected with LFQPOs because the observable frequencies obtained from the numerical results generally fall within this range seen in Table \ref{qpo_observational_counterparts}. In black-hole X-ray binaries, LFQPOs generally cover the frequency region below $30\,\mathrm{Hz}$ and are commonly used as tracers of accretion-state transitions \cite{Ingram:2019mna}. In the SKBH1 model, strong low-frequency components are found around $2.96\,\mathrm{Hz}$ and $6.62\,\mathrm{Hz}$, while the SKBH2 model has remarkable components around $3.86\,\mathrm{Hz}$ and $7.82\,\mathrm{Hz}$. These frequencies directly fall within the LFQPO range observed in Galactic black-hole binaries. Thus, SKBH1 and SKBH2 are better candidates than the Kerr reference model for interpreting slowly varying, coherent accretion-state oscillations.

Among the weak disformal deviation cases, the SKBH2 model appears as the most observationally interesting case. The pair formed at $3.86\,\mathrm{Hz}$ and $7.82\,\mathrm{Hz}$ appears as a strong low-frequency component, while at the same time this pair represents a more coherent higher component. This behavior resembles the observed tendency of black-hole binaries to show a dominant low-frequency QPO together with harmonic or secondary peaks. In this context, the SKBH2 model does not need to reproduce the exact $3:2$ high-frequency QPO ratio to be useful. The importance of this model is that a positive disformal deviation produces a stable, narrow, and observable QPO-like pattern in the same frequency range. The frequencies observed in this range are generally LFQPOs. This makes SKBH2 the best model in the present set for explaining persistent low-frequency QPO behavior in stellar-mass black-hole binaries.

Among the numerically calculated models, SKBH4 is not a very suitable model for clear twin-peak QPO systems. However, this model is much more suitable for broad-band noise and multiple transient peaks. The numerically calculated more coherent frequencies at $10.52\,\mathrm{Hz}$, $14.44\,\mathrm{Hz}$, and $29.31\,\mathrm{Hz}$ lie within the observable LFQPO range. However, when the overall PSD behavior is considered, it is more complex and less stable. For these reasons, the SKBH4 model is more suitable for explaining irregular or burst-like timing states. Observationally, this type of behavior is more appropriate for explaining QPO systems with strong broad-band variability rather than isolated narrow peaks.

The SKBH5 model is especially important for state-dependent timing behavior. In the early-time PSD analysis of this model, broad and weakly coherent peaks appear. In contrast, in the late-time phase, a strong and coherent peak forms at $4.58\,\mathrm{Hz}$. This behavior can be related to observations because observed black-hole binaries generally produce QPOs that are strongly connected to their accretion states. Thus, the SKBH5 model can be interpreted as a model in which a violent transition phase produces a much more coherent QPO-producing phase. Its late-time $4.58\,\mathrm{Hz}$ peak is observationally more meaningful than the broad early-time peaks, because it corresponds to the phase where the flow has reached a recurrent oscillatory configuration.

The mass-scaling formulas given in Eqs. \ref{fobs} and \ref{MBH} allow us to compare the numerical results with ULXs. The source M82 X-1 is especially important because the observed stable twin QPOs at $3.3\,\mathrm{Hz}$ and $5.1\,\mathrm{Hz}$ produce a value close to the $3:2$ ratio. These frequencies have been reported as an important indicator for revealing the presence of intermediate-mass black holes \cite{Pasham:2014ybe}. If the Kerr pair at $42.99\,\mathrm{Hz}$ and $68.13\,\mathrm{Hz}$ is scaled to the M82 X-1 pair, the implied black-hole mass is of order $100M_{\odot}$. This places the source M82 X-1 in the lower intermediate-mass range. At the same time, it shows that the same simulated hydrodynamical mechanism can explain observed sources by applying mass scaling from Galactic binaries to ULX systems. In addition, when the SKBH2 frequencies $3.86\,\mathrm{Hz}$ and $7.82\,\mathrm{Hz}$ are scaled to the observed M82 X-1 frequencies $3.3\,\mathrm{Hz}$ and $5.1\,\mathrm{Hz}$, the corresponding black-hole masses are approximately $11.7M_{\odot}$ and $15.3M_{\odot}$, respectively. Therefore, unlike the Kerr high-frequency pair, the SKBH2 low-frequency pair does not imply an intermediate-mass black hole for M82 X-1, but instead corresponds to a stellar-mass black-hole range under this direct frequency scaling.

For low-frequency ULX QPOs, the disformal models are much more useful. The source NGC 5408 X-1 shows a strong QPO near $20\,\mathrm{mHz}$, with evidence for a second feature near $15\,\mathrm{mHz}$ \cite{Strohmayer:2007pk}. When the numerically obtained $3.86\,\mathrm{Hz}$ frequency in the SKBH2 model is rescaled by using Eqs. \ref{fobs} and \ref{MBH}, the corresponding observed frequency becomes $20\,\mathrm{mHz}$, and the black-hole mass is calculated to be approximately $M=1900M_{\odot}$. Similarly, if the numerical QPO frequency of $4.58\,\mathrm{Hz}$ in the SKBH5 model is rescaled, the mass of the central black hole becomes approximately $M=2300M_{\odot}$. Thus, SKBH2 and SKBH5 provide plausible hydrodynamical counterparts for ULX mHz QPOs if the central object is in the intermediate-mass range.

The same argument can also be applied to massive black holes. The source RE J1034+396 is one of the best-known AGN QPO candidates. Its observed frequency is approximately $0.00027\,\mathrm{Hz}$. The period corresponding to this QPO is approximately $1$ hour \cite{Middleton:2008fe}. If the late-time peak frequency of $4.58\,\mathrm{Hz}$ numerically obtained in the SKBH5 model is scaled to this frequency, the mass of the central black hole is approximately $M=170000M_{\odot}$. On the other hand, when the peak frequency of $68.13\,\mathrm{Hz}$ found in the Kerr model is scaled to the observed frequency, the mass of the black hole in RE J1034+396 becomes approximately $M=2.5\times10^6M_{\odot}$. Both values fall in the mass range relevant for narrow-line Seyfert 1 galaxies, depending on whether the observed AGN oscillation is interpreted as the analogue of a low-frequency or high-frequency QPO.

Thus, as seen in  Table \ref{qpo_observational_counterparts}, the comparison between observational results and numerical results shows that different numerical models can explain different observational classes. The Kerr model appears as the best model for explaining HFQPOs such as the $41$--$67\,\mathrm{Hz}$ features observed from the source GRS 1915+105, while the SKBH1 and SKBH2 models appear as the most suitable models for LFQPOs in stellar-mass black-hole binaries. Here, the SKBH2 model produces the clearest coherent low-frequency pattern. The SKBH4 model can be interpreted as the best model for explaining observations in which irregular, multi-peak, burst-like timing behavior appears. The SKBH5 model is the most suitable model for state-dependent systems. In this model, after the transition phase, a phase in which coherent QPOs are produced emerges. Thus, the disformal spacetime parameters do not simply shift the QPO frequency; they select which observational timing class the simulated BHL accretion flow resembles.

\section{Conclusion}
\label{Concl}
In this work, we reveal the dynamical structure and observational signatures produced by BHL accretion around a slowly rotating disformal Kerr black hole. By numerically solving the GRH equations on the equatorial plane, we analyze how the disformal spacetime parameters modify the time-dependent behavior, stability, and morphological structure of the matter falling supersonically toward the black hole. The main aim of this study is not only to model the accretion dynamics, but also to reveal whether the resulting hydrodynamical variability carries QPO-like signatures. Thus, by connecting these results with observed timing features, it becomes possible to better understand black-hole systems. For this reason, in this work, we analyze two-dimensional density maps, one-dimensional azimuthal profiles, the time evolution of the mass accretion rate, PSD analysis, Lorentzian decomposition, and mass-scaled observational comparisons.

As a result of the numerical modeling, it has been shown that even small deviations from the Kerr geometry significantly modify the morphology of BHL accretion. In the SKBH1 model, where a weak negative deviation is applied, the shock cone still continues to exist. However, the opening angle of the cone becomes wider, and its azimuthal oscillation becomes stronger. In the SKBH2 model, where the deviation is weak but positive, the shock cone is still observed, but the shock-cone structure in the downstream region becomes more asymmetric and the density enhancement becomes more pronounced. These results show that weak disformal corrections are sufficient to modify the compression in the post-shock region, the angular spreading, and the oscillatory motion.

In the case of strong deviations, it is observed that the accretion dynamics changes dramatically. In the SKBH4 model, where the disformal combination is significantly smaller than the Kerr value, the classical shock cone is destroyed and replaced by a strongly non-axisymmetric mixed shock-cone/spiral structure. The matter in the post-shock region is redistributed around the black hole. As a result, irregular and burst-like accretion behavior is produced. In contrast, in the SKBH5 model, which has a strong positive deviation with respect to the Kerr model, a warped and shifted shock-cone structure is observed. In this case, the shock cone is not completely destroyed. Instead, it oscillates around the black hole and forms a recurrent late-time hydrodynamical state. Thus, the sign and strength of the disformal deviation control whether the flow behaves as a broadened shock cone, a distorted spiral structure, or a warped oscillating post-shock region.

The one-dimensional azimuthal profiles provide further support for these conclusions. The one-dimensional profiles of the density, radial velocity, and angular velocity show that the spacetime parameters do not only change the visual morphology of the flow, but also cause significant changes in the internal structure of the shock region. In particular, compared with the Kerr solution, the disformal models show a significant enhancement of the density in the post-shock region, changes in the radial velocity transition, and strong modifications in the discontinuities of the angular velocity. These changes show that, in the strong gravitational field, the disformal geometry directly affects the compression of the matter, the structure of the radial infall toward the black hole, and the azimuthal distribution of the matter in the post-shock region.

The mass accretion-rate analysis shows that the morphological differences formed around the slowly rotating disformal Kerr black hole are clearly seen in the time-dependent behavior of the mass accretion rate. In the Kerr model, an almost steady accretion rate is formed, although small oscillations are also observed. However, in the disformal models, significant variations in the mass accretion rate are revealed depending on the spacetime parameters and the deviation. While moderate oscillations are observed in the SKBH1 model, it is shown that SKBH2 produces stronger and persistent quasi-periodic modulations. On the other hand, in the SKBH4 model, irregular, intermittent, burst-type accretion is formed due to the strongly distorted mixed shock-cone/spiral morphology. The SKBH5 model produces two different temporal regions: an early violent transition phase and a later coherent oscillation phase. Thus, the accretion-rate signals act as a bridge that reveals the connection between the hydrodynamical structure of the flow and the observed QPO-like timing signatures.

From the PSD and Lorentzian analyses, it is numerically observed that each spacetime configuration produces a different frequency-domain signature. In the Kerr reference model, coherent peaks are numerically obtained around $42.99\,\mathrm{Hz}$ and $68.13\,\mathrm{Hz}$, and these frequencies are found to be consistent with the HFQPOs observed from the source GRS 1915+105. In the SKBH1 model, the deviation from Kerr shows that the timing power in the PSD shifts toward lower frequencies. In this case, the observable coherent frequencies are calculated to occur around $2.96\,\mathrm{Hz}$ and $6.62\,\mathrm{Hz}$. In the SKBH2 model, clear and highly coherent LFQPO behavior is observed, especially at $3.86\,\mathrm{Hz}$ and $7.82\,\mathrm{Hz}$. The numerical frequencies obtained in the SKBH2 model show that this model appears as the most remarkable weak-deviation model for LFQPO sources observed in the literature. The SKBH4 model produces a more complex PSD with multiple peaks embedded in a broad and turbulent signal, making it more suitable for irregular or burst-like timing states. In the SKBH5 model, after the transition phase, a strong late-time coherent peak is observed around $4.58\,\mathrm{Hz}$. This model is especially important for explaining state-dependent QPO behavior.

As a result of the comparison between the numerical results and observations, it has been shown that different models can explain QPOs obtained from different black-hole sources observed in the literature. In addition, by comparing the observational results with the numerical results, a prediction has been made about what the black-hole mass may be for some sources. It has been shown that the Kerr model is the best model for explaining HFQPOs. This is because the numerically calculated QPO frequencies are found to be consistent with the $41$--$67\,\mathrm{Hz}$ range observed from the source GRS 1915+105. The weak disformal models SKBH1 and SKBH2 naturally show similarities with the LFQPOs observed in Galactic black-hole binaries. In particular, the numerical frequencies calculated in the SKBH2 model are found to produce a clear coherent low-frequency pattern. The SKBH4 model is found to be a model that can explain observed sources showing broad-band noise, multiple transient peaks, and irregular timing behavior. The SKBH5 model is found to be highly consistent with observed sources in which QPOs depend on the accretion state, because it naturally produces a transition from a violent early phase to a coherent late-time QPO-producing phase.

Finally, inverse-mass scaling is used to extend the frequencies numerically calculated for a black-hole mass of $M=10M_{\odot}$ to intermediate-mass and massive black holes. In this way, when the frequencies obtained numerically in different models, ranging from a few Hz to a few tens of Hz, are shifted to sources with different black-hole masses, the resulting frequencies are found to vary between the mHz and micro-Hz ranges. This allows the numerical results to be compared with ULX sources such as M82 X-1 and NGC 5408 X-1, as well as AGN QPO candidates such as RE J1034+396. Thus, the disformal spacetime parameters do not only cause the QPO frequencies to shift, but also define the hydrodynamical region of the accretion flow and allow us to determine which observational timing class is consistent with the numerically obtained hydrodynamical mechanisms. This makes BHL accretion in disformal Kerr geometry a physically powerful framework for connecting modified-gravity spacetime structure with observable black-hole timing signatures.


\section*{Acknowledgments}
All numerical simulations were performed using the Phoenix High
Performance Computing facility at the American University of the Middle East (AUM), Kuwait.

\section*{Data Availability Statement}
The data sets generated and analyzed during the current study were produced using high-performance computing resources. These data are not publicly available due to their large size and computational nature, but are available from the corresponding author upon reasonable request.

\bibliographystyle{apsrev4-2}
\bibliography{reference}

\end{document}